\documentclass[twocolumns]{emulateapj}

\usepackage{amsmath, amssymb}
\usepackage{graphicx,color,colordvi}
\usepackage{latexsym}
\usepackage{url}
\usepackage{times}
\usepackage{longtable}

\def \bmath #1 {{\hbox{\boldmath{$#1$}\unboldmath}}}

\slugcomment{Submitted to Astrophys. J.}

\shorttitle{OH$^+$: state-to-state formation, 
Einstein coefficients and inelastic rates with He}

\shortauthors{G\'omez-Carrasco  et al.}

\def \HH{\ifmmode{\rm H_2}\else{$\rm H_2$}\fi}
\def \Cp{\ifmmode{\rm C^+}\else{$\rm C^+$}\fi} 
\def \Cpp{\ifmmode{\rm C^{++}}\else{$\rm C^{++}$}\fi} 
\def \CHp{\ifmmode{\rm CH^+}\else{$\rm CH^+$}\fi}
\def \CHdp{\ifmmode{\rm CH_2^+}\else{$\rm CH_2^+$}\fi}
\def \CHtp{\ifmmode{\rm CH_3^+}\else{$\rm CH_3^+$}\fi} 
\def \HdO{\ifmmode{\rm H_2O}\else{$\rm H_2O$}\fi} 
\begin{document}

   \title{OH$^+$ in astrophysical media: state-to-state formation rates, Einstein coefficients and inelastic collision rates with He
    }

   \author{ Susana G\'omez-Carrasco $^a$,  Benjamin Godard$^b$, Fran\c{c}ois Lique$^c$,  Niyazi Bulut$^d$, 
            Jacek K{\l}os$^e$,
            Octavio Roncero$^{f,1}$, Alfredo Aguado$^g$, F. Javier Aoiz$^h$, Jes\'us F. Castillo$^h$, 
            Javier R. Goicoechea$^i$, Mireya Etxaluze$^i$  and Jos\'e Cernicharo$^i$.
   }\footnote{e-mail: octavio.roncero@csic.es}

       \affil{ (a) Facultad de Qu{\'\i}mica, Unidad Asociada CSIC-USAL,  Universidad de Salamanca, Spain. }

       \affil{ (b) LERMA, CNRS UMR 8112, Observatoire de Paris, Meudon, France}

        \affil{ (c) LOMC - UMR 6294, CNRS-Universit\'e du Havre, 25 rue Philippe Lebon, BP 540, 76058,
                        Le Havre, France}

       \affil{ (d) Department of Physics, Firat University, 23169 Elazi{\~g}, Turkey}

        \affil{ (e) Department of Chemistry and Biochemistry, University of Maryland, College Park, MD 20742-2021, USA}
                          
       \affil{ (f)  Instituto de F{\'\i}sica Fundamental (IFF-CSIC),  C.S.I.C., 
                       Serrano 123, 28006 Madrid, Spain. }
                          
       \affil{ (g)  Facultad de Ciencias,  Unidad Asociada de Qu{\'\i}mica-F {\'\i}sica Aplicada CSIC-UAM, 
                          Universidad Aut\'onoma de Madrid, Spain}
                          
       \affil{ (h)  Departamento de Qu{\'i}mica F{\'i}sica I, Unidad Asociada de Qu{\'\i}mica-F{\'\i}sica CSIC-UCM,
                         Facultad de Qu{\'\i}mica, Universidad Complutense de Madrid, Spain}
                                         
       \affil{ (i) Instituto de Ciencia de Materiales (ICMM-CSIC), C.S.I.C., 
                          Sor Juana In\'es de la Cruz, 3, Cantoblanco, 28049 Madrid, Spain}

     \begin{abstract}
The rate constants required to model the OH$^+$ observations in different regions of the 
interstellar medium have been determined using  state of the art quantum methods.
 First, state-to-state rate constants for the H$_2(v=0,J=0,1)$+ O$^+$($^4S$) 
$\rightarrow$ H + OH$^+(X ^3\Sigma^-, v', N)$ reaction have been obtained using a quantum wave packet 
method. The calculations have been compared with time-independent results to asses the accuracy 
of reaction probabilities at collision energies of about 1 meV. The good agreement
between the simulations and  the existing 
experimental cross sections in the $0.01-$1 eV energy range shows the quality of the results. 
 The calculated state-to-state rate constants have been fitted 
to an analytical form. Second, the Einstein coefficients of OH$^+$
have been obtained for all astronomically significant ro-vibrational bands
involving the   $X^3\Sigma^-$ and/or  $A^3\Pi$ electronic states.
 For this purpose the potential energy curves and electric 
dipole transition moments for seven electronic states of OH$^+$ are calculated with {\it ab initio} 
methods at the highest level and including spin-orbit terms, and the rovibrational levels 
have been calculated including the empirical spin-rotation and spin-spin terms. Third, the 
state-to-state rate constants for inelastic collisions between He and OH$^+(X ^3\Sigma^-)$ have been 
calculated using a time-independent close coupling method on a new potential energy surface.
All these rates have been implemented in detailed chemical and radiative transfer models.
Applications of these models to  various astronomical sources show that inelastic collisions dominate the excitation 
of the rotational levels of OH$^+$. In the models considered
 the excitation resulting from the  chemical formation of OH$^+$ increases 
the line fluxes by about 10 \% or less depending on the density of the gas.
     \end{abstract}
       \centerline{\today}

\maketitle

%%%%%%%%%%%%%%%%%%%%%%%%%%%%%%%%%%
\section{Introduction}
%%%%%%%%%%%%%%%%%%%%%%%%%%%%%%%%%%

% ########################################################################################
Light hydrides represent the very first step of interstellar chemistry. They start reaction cycles 
that initiate the formation of complex molecules and are therefore at the root of the molecular 
richness observed for decades in all interstellar environments. In addition, because of the 
diversity of their formation and excitation pathways, their rotational lines offer powerful 
diagnostics of  the physical and chemical processes at play in the interstellar medium (ISM).

These investigations have recently been deepened by the Herschel satellite which opened
 the spectral domain of hydrides absorption and emission that was
 not accessible to us before due to  the large opacity of the Earth
atmosphere. Indeed, many light hydrides (e.g. CH, CH$^+$, HF, HCl, OH$^+$, H$_2$O, NH, SH$^+$) 
have been observed, some of them for the first time, in different types of interstellar and 
circumstellar regions (e.g. \citealt{Benz2010,Cernicharo2010,Hily-Blant2010,
Gerin-etal:10,Naylor2010,van-Dishoeck2011,Neufeld2011,Godard2012,Spoon2013}).

Among all the hydrides detected to date, the hydroxyl cation OH$^+$ is particularly interesting, 
not only because it initiates the oxygen chemistry, but also because its abundance could be a 
valuable tracer of cosmic ray and X-ray ionization rates \citep{Gerin-etal:10,Hollenbach-etal:12,
Gonzalez-Alfonso-etal:13}. First detected in absorption in the diffuse medium 
\citep{Wyrowski2010,Gerin-etal:10,Neufeld-etal:10,Krelowski-etal:10,Porras-etal:14}, OH$^+$ 
has then been also observed in absorption and emission in a variety of interstellar and circumstellar
environments 
including hot and dense photodissociation regions (PDR) \citep{vanderTak-etal:13,Pilleri-etal:14}, 
galactic center clouds \citep{Etxaluze-etal:13,Goicoechea-etal:13}, planetary nebulae
\citep{Etxaluze-etal:14,Aleman-etal:14} and the nuclei 
of active galaxies \citep{van-der-Werf2010,Gonzalez-Alfonso-etal:13} . These observations need to be interpreted
both in terms of chemistry and excitation processes for which important properties of the OH$^+$ molecule
are still lacking.

The chemistry of OH$^+$ in molecular clouds is rather well understood.
As the depth into the cloud increases and the far ultraviolet (FUV) flux decreases, the formation of OH$^+$ 
successively follows two different pathways \citep{Hollenbach-etal:12}. At the border of 
diffuse clouds where most of the Hydrogen is atomic, the production of OH$^+$ proceeds through
\begin{eqnarray}\label{ion+H2-reaction}
{\rm H} \xrightarrow{\rm CR} {\rm H}^+ \xrightarrow{\rm O} {\rm O}^+ \xrightarrow{\rm H_2} {\rm OH}^+,
\end{eqnarray}
being initiated by the ionization of atomic Hydrogen by cosmic rays (CR).
Conversely, deeper in the cloud where most of the Hydrogen is molecular, the production 
of OH$^+$ proceeds through
\begin{eqnarray}\label{ion+H2-reaction}
{\rm H}_2 \xrightarrow{\rm CR} {\rm H}_2^+ \xrightarrow{\rm H_2} {\rm H}_3^+ \xrightarrow{\rm O} {\rm OH}^+.
\end{eqnarray}
As a result, the abundance of OH$^+$ predicted by chemical models displays 
two peaks whose position and magnitude depend on the ratio of the incident UV radiation field
and the gas density \citep{Hollenbach-etal:12}.

On the other hand and because of the lack of theoretical and experimental data, uncertainties
remain on the processes involved in populating the rotational levels of OH$^+$. Firstly, 
these levels might be excited by inelastic collisions. In the cold ISM, the most abundant 
species are H$_2$ and He. In warmer regions such as diffuse, translucent clouds or PDRs, 
collisions with electrons and atomic Hydrogen should also be taken into account. Secondly, 
since OH$^+$ is observed in hot PDRs illuminated by strong infrared (IR) and UV radiation 
fields, its rotational levels might be sensitive to the radiative pumping of its vibrational 
and electronic states followed by radiative decay. Indeed, these mechanisms have been found 
to dominate the excitation of many species (e.g. H$_2$O, HNC, NH$_3$, H$_2$, CO, CH$^+$) in molecular 
clouds and circumstellar envelopes (e.g. \citealt{Gonzalez-Alfonso1999,Agundez-Cernicharo:06,Agundez2008,Troutman2011,
godard:13}). At last, since OH$^+$ is a very reactive molecule it was assumed
that it is destroyed before inelastic collisions may take place and it has been proposed that 
its rotational population is governed  by its chemical formation \citep{vanderTak-etal:13}. 
For instance, it has recently been shown that  chemical state-to-state formation
pumping plays a major role in 
the  excitation of several molecules such as CH$^+$ in hot and dense PDRs, planetary nebulae 
and circumstellar disks \citep{Nagy2013,godard:13,Zanchet-etal:13,Zanchet-etal:13b}.

The purpose of this paper is to provide the excitation rates of OH$^+$ through radiative 
pumping, inelastic collisions, and reactive collisions in order to improve the reliability 
of chemical and radiative transfer models applied to astrophysical environments. It is 
organized as follows: in Sect. II, we study the excitation of OH$^+$ during its chemical formation via O$^+(^4S)$ 
+ H$_2(v,J)$  $\rightarrow$ OH$^+(X ^3\Sigma^-,v',N )$ + H. Up to now, only total reaction cross 
sections and rate constants have been obtained, both experimentally \citep{Burley-etal:87}
and theoretically using quasi-classical trajectory (QCT) calculations \citep{Martinez-etal:05}, 
time independent calculations with hyperspherical coordinates (TI), and wave packet (WP) 
methods \citep{Martinez-etal:06b,Xu-etal:12} using the adiabatic ground electronic 
state potential energy surface (PES) of \cite{Martinez-etal:04};
in Sect. III, we focus on the radiative pumping of the vibrational and electronic levels
of OH$^+$ by infrared and UV photons. In a previous study, \cite{Almeida-Singh:81} have 
reported the absolute oscillator strengths for several vibrational states of the OH$^+$ 
$(X^3\Sigma^-, A^3\Pi)$ system based on the radiative lifetimes measured by \cite{Brzozowski-etal:74}
and the ultraviolet  emission  spectra of OH$^+$ observed by \cite{Merer-etal:75}.
To extend these results to higher $J$ and $v$ values, we compute here the {\it ab initio} 
potential energy curves of the OH$^+(X ^3\Sigma^-, A^3\Pi)$ band system;
in Sect. IV, we investigate the excitation of OH$^+$ by inelastic collisions with He and
use the results as a model for collisions with H and H$_2$. So far, previous studies have 
only reported excitation rates by electron impact \citep{schoier:05,vanderTak-etal:13}. We compute here the 
interaction potential of the He-OH$^+ (X ^3\Sigma^-)$ system and perform scattering calculations 
in order to derive for the first time the associated collisional rate constants;
finally, we discuss, in Sects. V and VI, the implication of all these results on the modeling 
of astrophysical environments and the interpretation of observations.

For reasons of clarity and comprehensibility of the manuscript, the details of the reactive collisions
are given in Appendix A together with the list of parameters obtained to fit the state-to-state rate constants; 
the {\it ab initio} details for the calculation of the 7 electronic states of OH$^+$ and 
the Einstein coefficients are described in Appendix B; the calculation of the  PES built
in this work  for the He + OH$^+(X ^3\Sigma^-)$ is described in Appendix C; finally, the time-independent
calculation details of the  He + OH$^+(X \Sigma^-)$ inelastic collisions are given in Appendix D.

%%%%%%%%%%%%%%%%%%%%%%%%%%%%%%%%%%%%%%%%
\section{Reactive collision simulations}
%%%%%%%%%%%%%%%%%%%%%%%%%%%%%%%%%%%%%%%%

The state-to-state rate constants for the  O$^+(^4S)$ 
+ H$_2(v,J)$  $\rightarrow$ OH$^+(X ^3\Sigma^-)$ + H reaction
have been calculated using a time dependent WP method (find further
details in Appendix A) on the ground electronic 
state PES of \cite{Martinez-etal:04}. The details
of the computations are described in Appendix A together
with the convergence analysis and comparison with  results
obtained using time-independent methods.

 OH$^+(X ^3\Sigma^-)$ products are treated in Hund's case (b), and the total diatomic angular momentum 
is ${\bf J}={\bf N}+{\bf S}$, with ${\bf N}$ being
the total rotational angular momentum and ${\bf S}$ the total electronic spin.
 In the simulation of the  reactive collision rates, the effect of the electronic spin 
is neglected. Under this approximation, several alternatives are possible 
to determine the population of the three $F_1(J=N+1)$, $F_2(J=N)$ and $F_3(J=N-1)$
levels. One possibility would be to consider the three levels equally
populated. However, we shall assume here that the population 
of each $F_i$ sublevels is proportional to the degeneracy $(2J+1)$.

%%%%%%%%%%%%%%%%%%%%%%%%%%%%%%%%%%%%%%%%%%%%%%%%%%%%%%%%%%%%%%
\begin{figure}[t]
\vspace*{-0.0cm}

\hspace*{0.5cm}
\includegraphics[scale=0.45,angle=0]{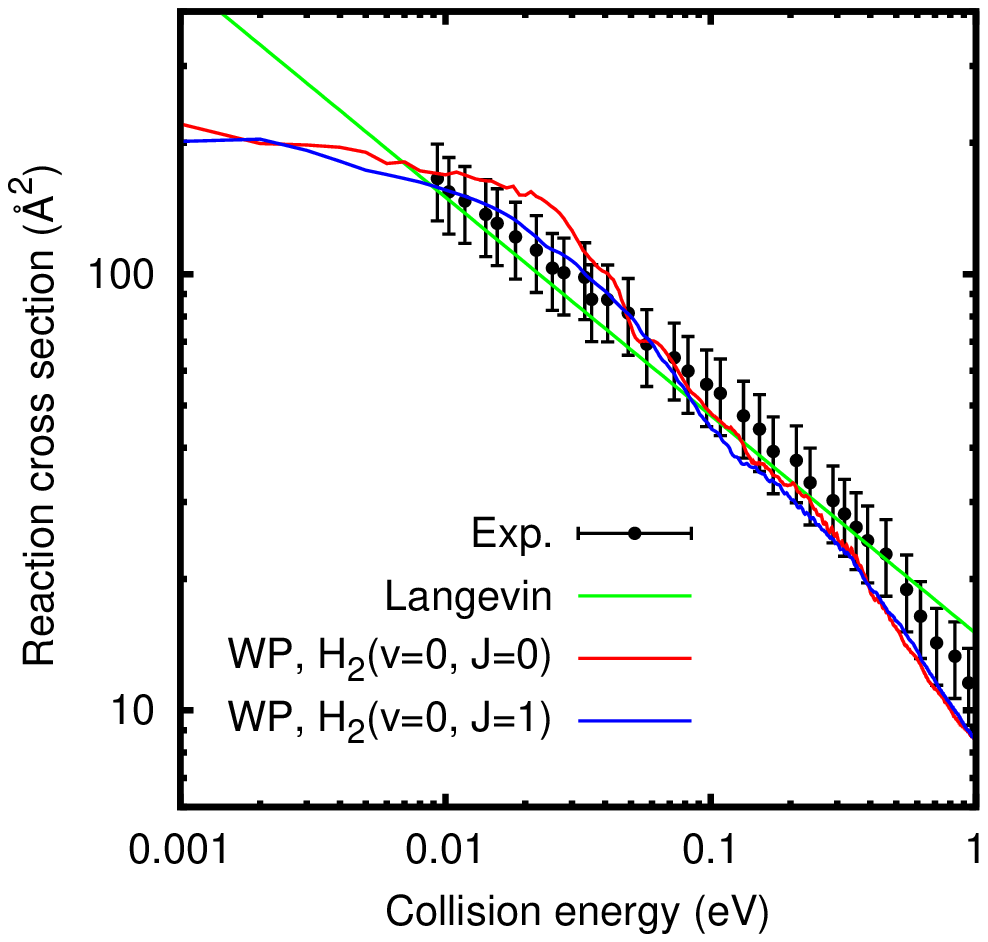}
\vspace*{-0.0cm}
   \caption{\label{ICS}
Total reaction cross section 
for the H$_2(v,=0,J=0,1)$+O$^+$ collisions
as a function of collision energy.
The experimental results are taken from \cite{Burley-etal:87}.
The Langevin results come from the typical Lagevin
model $\sigma(E)=A /\sqrt{E}$, with $A$=15 \AA$^2$ eV$^{1/2}$.
}
\end{figure}
%%%%%%%%%%%%%%%%%%%%%%%%%%%%%%%%%%%%%%%%%%%%%%%%%%%%%%%%%%%%%%%%%

 The total reaction integral  cross section (ICS) is  obtained after partial wave summation
as described in Appendix A, and is compared in Fig.~\ref{ICS}
with both the experimental results of \cite{Burley-etal:87} and the results
obtained using the Langevin model \citep{Langevin:1905,Gioumousis-Stevenson:58}.
The Langevin model is in good agreement with the experimental results, although slightly 
lower for collision energies $< 0.5$ eV. \cite{Burley-etal:87} attributed this small difference
to the simplicity of the Langevin model, which only takes into account the distance 
between reactants, but not the anisotropy of the reaction. 

The ICS results for $J = 0$ and 1 are very close to each other, except for energies 
below 0.04 eV where the cross section for $J=0$ is slightly higher. The agreement with the 
experimental data is very good, nearly always inside the experimental error bars.
It is interesting that for $E<0.01$ eV the calculated cross section deviates
from the simple Langevin model.  These small inaccuracies affect more notoriously to low 
collision energies. For energies above 0.3-0.4 eV, the WP results become slightly below 
the experimental error bars, probably due to small inaccuracies of the PES.
The experimental results also show a change of the energy dependence at these energies 
with respect to the pure Langevin model. It can be concluded that the simulated cross 
sections are in very good agreement with the available experimental results, and can 
thus be used to estimate rate constants. 

%%%%%%%%%%%%%%%%%%%%%%%%%%%%%%%%%%%%%%%%%%%%%%%
\begin{figure}[b]
\vspace*{-0cm}

\hspace*{-0.2cm}
\includegraphics[scale=0.5,angle=0]{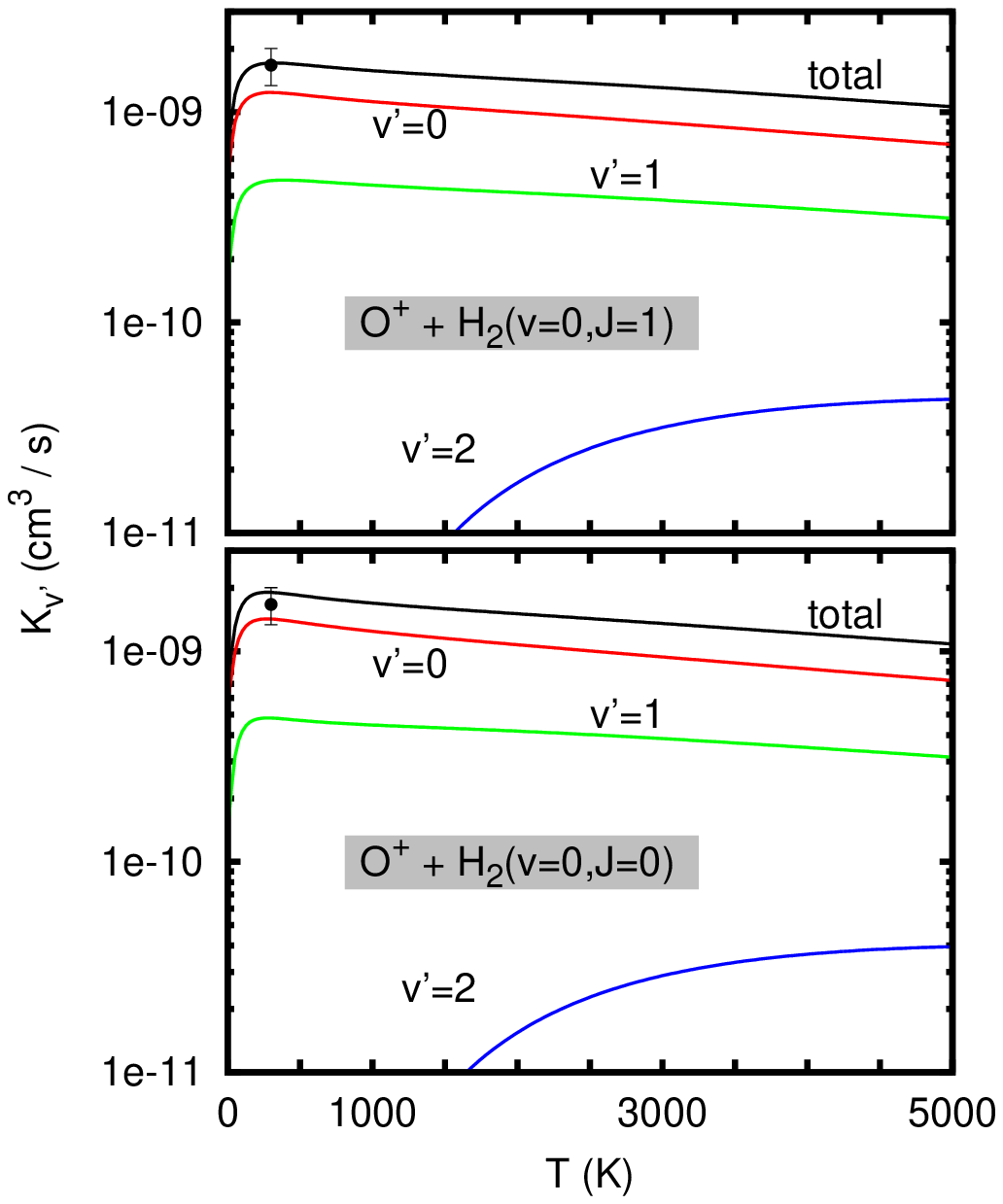}

\vspace*{-0.0cm}
   \caption{\label{ratevib}
Total and vibrationally resolved rate
constants for the O$^+$ +H$_2(v,=0,J=0,1)$ collisions
as a function of translational temperature.
The experimental result of 1.67 10$^{-9}$ cm$^3$/s  is taken from \cite{Burley-etal:87}.
}
\end{figure}
%%%%%%%%%%%%%%%%%%%%%%%%%%%%%%%%%%%%%%%%%%%%%

The rate constants in the 50-5000 K temperature range
 are obtained by numerical integration of the state-to-state cross sections,
in the 1meV-1.5eV energy range . 
 The total and vibrationally 
resolved rate constants are shown in Fig.~\ref{ratevib}. The agreement with the only experimental
value of \citet{Burley-etal:87}  is good.  The rates for $v'=0$ and 1 are both significant
over the whole range of temperatures considered, while the rates to  OH$^+$ products in 
$v' > 1$ are negligible. The results derived for H$_2$($v=0,J=0,1$) are within the error bars
of the experimental value obtained by \citet{Burley-etal:87} at room temperature. The total 
rates show a slight decrease with increasing temperature. This departure from a pure Langevin 
behavior is due to the long range behaviour of the PES, which is not isotropic 
as assumed in the simple mono-dimensional Langevin model.

%%%%%%%%%%%%%%%%%%%%%%%%%%%%%%%%%%%%%%%%%%%%%%%%%%%%%%%%%%%%%%%%%%
\begin{figure}[t]

\vspace*{-.cm}

\hspace*{0.1cm}\includegraphics[scale=0.47,angle=0]{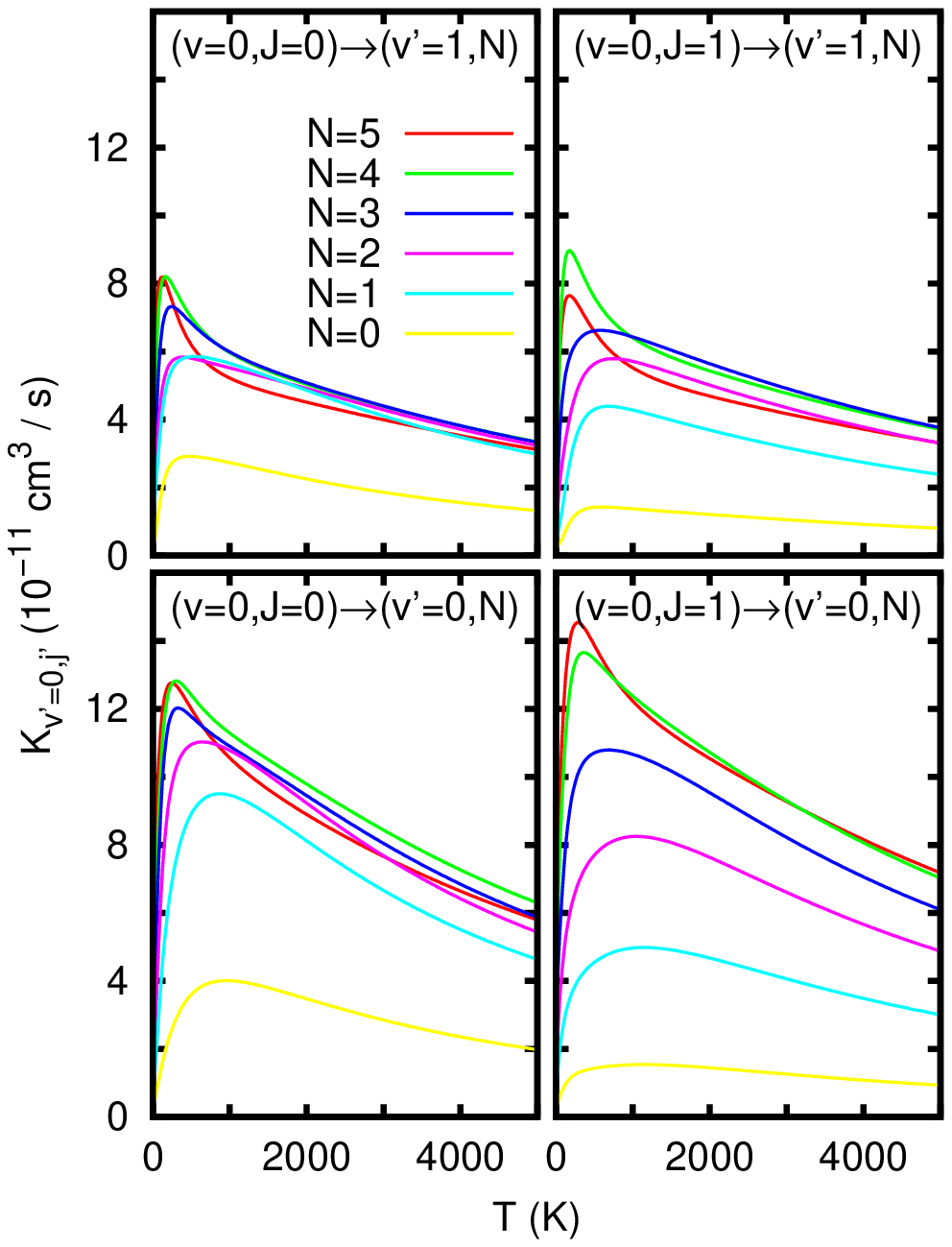}

\vspace*{-.0cm}
   \caption{\label{raterot}
State-to-state rate
constants for the H$_2(v,=0,J=0,1)$+O$^+\rightarrow$ OH$^+(X ^3\Sigma^-, v'=0,1,N)$ + H collisions
as a function of translational temperature, with the quantum numbers being specified
in each panel. 
}
\end{figure}
%%%%%%%%%%%%%%%%%%%%%%%%%%%%%%%%%%%%%%%%%%%%%%%%%%%%%%%%%%%%

The rotationally resolved state-to-state rate constants are shown in Fig.~\ref{raterot}, for 
initial rotational states of H$_2$, $J = 0$ (left panels) and $J =1$ (right panels), and final 
vibrational states of OH$^+$, $v' = 0$ (bottom panels) and $v' = 1$ (top panels). The rates 
increase with increasing $N$, reaching a maximum at about $N = 6$ or 7, indicating that reactive
collisions between O$^+$ and H$_2$ proceeds through a significant energy transfer in the 
excited states of the product OH$^+$. In fact, for initial H$_2(J=1)$ the final OH$^+(N)$ 
products seem to be more excited by just one rotational quantum. The rates for $v'=1$ are 
approximately 2/3 of those obtained for $v' = 0$ and show similar behaviors.

For $v'=0$ and $N< 16$, the reaction has no threshold. However, at low temperatures, the 
rates for low $N$ are nearly zero, increasing rapidly with temperature, reaching a maximum 
at about 300-800 K and decreasing again afterwards, so that at high temperatures the rates 
for all $N$ become rather similar. For $v'=0$ and $N>16$, there is an energy threshold and 
the rates increase monotonously with increasing temperature. In the case of $v'=1$, the 
behaviour is similar but the threshold appears at $N\approx 8$.

The state-to-state rate constants have been fitted as described in Appendix A
and the parameters obtained are listed
 in Table~\ref{tabla-v}.

%%%%%%%%%%%%%%%%%%%%%%%%%%%%%%%%%%%%%%%%%%%%%%%%%%%%%%%%%%%%%%%
\section{Einstein coefficients for OH$^+(X^3\Sigma^-, A^3\Pi)$}
%%%%%%%%%%%%%%%%%%%%%%%%%%%%%%%%%%%%%%%%%%%%%%%%%%%%%%%%%%%%%%%

In the previous section, the H$_2$ + O$^+(^4S)\rightarrow$ H+OH$^+(X ^3\Sigma^-)$ reaction 
has been studied in the ground adiabatic electronic state, without considering electronic 
spin.  In this section we study the transitions between 
the X and A electronic states of OH$^+$. To do so it is important to consider that the 
total angular momentum of OH$^+$ is ${\bf J}={\bf N}+ {\bf S}$, with ${\bf N}={\bf R}+ 
{\bf L}$ where ${\bf R}$ is the rotational angular momentum, and ${\bf L}$ and ${\bf S}$ 
are the electronic orbital and spin angular momenta, respectively. 

\begin{figure}[b]
\hspace*{-0cm}\includegraphics[scale=0.45,angle=0]{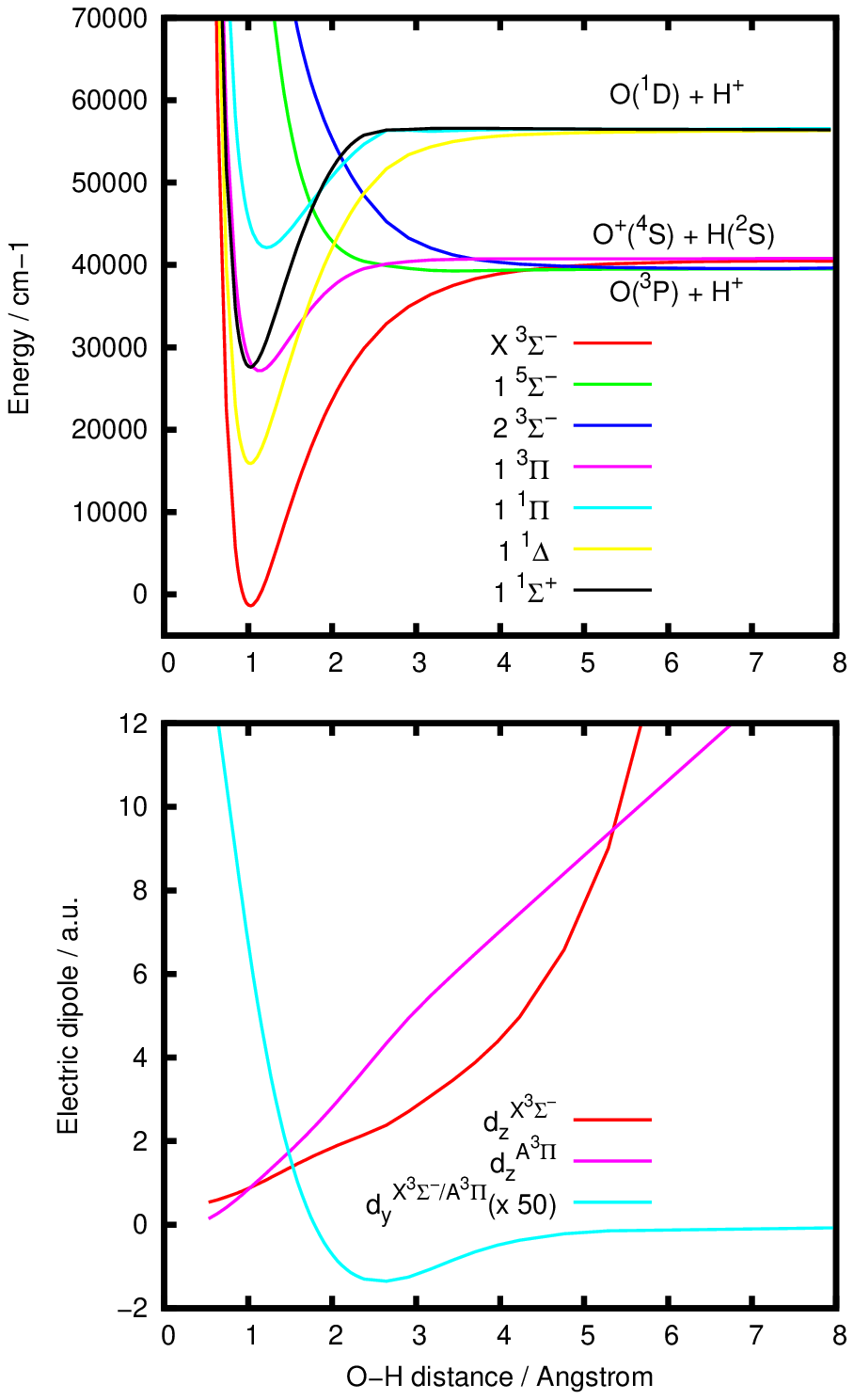}
\vspace*{-0.0cm}
\caption{ \label{OHp_electronic_states} Top: Electronic states of 
the OH$^{+}$ cation. 
Bottom: Dipole moments and transition dipole moments between some 
of the electronic states. These results are obtained using the
extrapolation up to complete basis set as explained in the text.
}
\end{figure}

Seven electronic states of OH$^+$ have been calculated to describe properly
the dissociation asymptotes of OH$^+(X^3\Sigma^-, A^3\Pi)$ as described in Appendix B.
The potential energy curves  are displayed in the top panel of Fig. 
\ref{OHp_electronic_states}. The dipole moments of the X$^3\Sigma^{-}$ ground state and 
the A$^3\Pi$ excited state along with the their transition dipole moments are displayed
in the bottom panel of Fig. \ref{OHp_electronic_states}. In the following we focus
on the X$^3\Sigma^{-}$ and $A^3\Pi$ states only.

%%%%%%%%%%%%%%%%%%%%%%%%%%%%%%%%%%%%%%%%
%%%%%%%%%%%%%%%%%%%%%%%%%%%%%%%%%%%%%%%%%

The ultraviolet $A^3\Pi-X^3\Sigma^-$ emission spectra of OH$^+$ was studied by \cite{Merer-etal:75}.
After including the spin-spin and spin-rotation terms and the $\Lambda$-doubling of the $A^3\Pi$ 
state due to the $A^1\Delta$ state, the deperturbed potential energy curves were obtained. The 
equilibrium distances for $X^3\Sigma^-$ and $A^3\Pi$ states in 
Table 8 of \cite{Merer-etal:75} are $R_e= 1.028$ and 1.135\AA\, respectively, 
in good agreement with the values of 1.0284 and 1.1356 \AA\ 
obtained here. \cite{Merer-etal:75} found a vertical excitation energy  
of $T_e= 28438.55$ cm$^{-1}$ while here a value of 28522.65 
cm$^{-1}$ is obtained. Finally, the dissociation energies obtained  here are 41900.0 and 
13661.8 cm$^{-1}$ for for $X^3\Sigma^-$ and $A^3\Pi$ states, respectively.

The details for the computation of the ro-vibrational states of OH$^+$ in the
 $X^3\Sigma^-$ and $A^3\Pi$ electronic states, and the  Einstein coefficients,
are described  in  Appendix B.

The radiative lifetimes of the $ A^3\Pi_\Omega, v,J $ states are obtained as the inverse 
of the of sum of all possible $A^3\Pi_\Omega, v,J \rightarrow X^3\Sigma^-, v',J'$ 
transitions \citep{Larsson1983}. Using the summation rule of the H\"onl-London factors 
\citep{Whiting-Nicholls:74,Whiting-etal:80} we define a vibrational lifetime (as done by 
\citealt{Larsson1983})
\begin{eqnarray}\label{radiative-lifetime}
\tau_v = \left( \sum_{v'} 
 {1 \over 3\pi\epsilon_0 \hbar^4} \left({ h {\overline \nu} \over c}\right)^3 
  \delta_{SS'}
 \left\vert M_{v;v'}^{J\Lambda S \alpha; J' \Lambda' S'\alpha'} \right\vert^2\right)^{-1},
\end{eqnarray}
where $h {\overline \nu}$ is the average transition energy
and the $M_{v;v'}^{J\Lambda S \alpha; J' \Lambda' S'\alpha'}$ are described in Appendix B.
 These radiative lifetimes 
are listed in Table~\ref{table-radiative-lifetimes} for OH$^+(A ^3\Pi,v)$.
There are two experimental studies reporting very different lifetimes for  OH$^+(A ^3\Pi,v=0)$. 
\cite{Brzozowski-etal:74} reported lifetimes for $v$ between 850 and 1010 ns, obtained by 
averaging over rotational bands for particular OH$^+(A ^3\Pi,v\rightarrow ^3\Sigma^-, v')$ 
bands. Later \cite{Mohlmann-etal:78} reported a radiative lifetime for  OH$^+(A ^3\Pi,v=0)$
of 2500 ns, in very good agreement with the results of the present work,
but considerably longer than that reported previously \citep{Brzozowski-etal:74}. 
These authors argued that this difference is originated from the effect of the pressure
on the lifetimes in the case of long-range interactions present
in charged gases. This makes necessary to carry out pressure dependent measurements 
to extrapolate to zero pressure to get reliable radiative lifetimes, as done 
by \cite{Mohlmann-etal:78}. We may therefore conclude, that our results are rather
reliable. It should be noted that the Einstein coefficients obtained by
\cite{Almeida-Singh:81} are based on the experimental radiative lifetimes 
of \cite{Brzozowski-etal:74}, and are therefore 2.5 times larger than those reported in
this work.

%%%%%%%%%%%%%%%%%%%%%%%%%%%%%%%%%%%%%%%%%%%%%%%%%%%%%%%%%%%%%%%%%%%%%
 \begin{table}[h]
\label{table-radiative-lifetimes}
 \caption{
\rm{Radiative lifetimes, $\tau_v$, for the vibrational
states of the OH$^+(A ^3\Pi,v)$ states calculated  using Eq.~(\ref{radiative-lifetime}).
}}
 \begin{center}
 \begin{tabular}{|c|c|}
 \hline 
 $v$ & $\tau_v$ (ns) \\
 \hline
    0 &  2524\\
    1 &  2665 \\
    2 &  2820\\
    3 &  3004\\
    4 &  3233\\
    5 &  3534\\
    6 &  3960\\
    7 &  4637 \\
    8 &  5961\\
    9 &  9559 \\
   10 &  16118\\ 
 \hline
 \end{tabular}
 \end{center}
 \end{table}
%%%%%%%%%%%%%%%%%%%%%%%%%%%%%%%%%%%%%%%%%%%%%%%%%%%%%%%%%%%%%

%%%%%%%%%%%%%%%%%%%%%%%%%%%%%%%%%%%%%%%%%%%%%%%
\section{Collisional excitation of OH$^+(X ^3\Sigma^-)$ by He}
%%%%%%%%%%%%%%%%%%%%%%%%%%%%%%%%%%%%%%%%%%%%%%%
In this section, we consider the collisional excitation
 of OH$^+$ by He.
Helium, as a closed shell atom with two electrons, is sometimes considered as a reasonable template
 of molecular Hydrogen 
\citep{schoier:05,lique08b}. However, for a molecular cation 
such as OH$^+$, such approximation is expected to be moderately accurate due 
to the fact that  the interaction of He and H$_2$ with an ion significantly differs.
 Generally, He rate coefficients underestimate H$_2$ rate coefficients by a factor 
that can be up to an order of magnitude \citep{Roueff:13}.

In addition, OH$^+$ can react with H$_2$ to form H$_2$O$^+$. Then, it is really 
quite uncertain to estimate H$_2$-rate constants from the He ones. Nevertheless, 
we expect that the present data will enable rough estimate of the collisional 
excitation process of OH$^+$ in the ISM that is crucial for modeling the abundance
 and excitation of the OH$^+$ molecule. 

To the best of our knowledge, no collisional rate constants for the OH$^+$ 
molecule have been published before.
Within the Born-Oppenheimer approximation, scattering cross sections and corresponding
 rate constants are obtained by solving  the motion of the nuclei on an electronic
 PES, which is independent of the masses and spins of the nuclei. 

\begin{figure}[t]
\vspace*{-0cm}
\includegraphics[width=8cm]{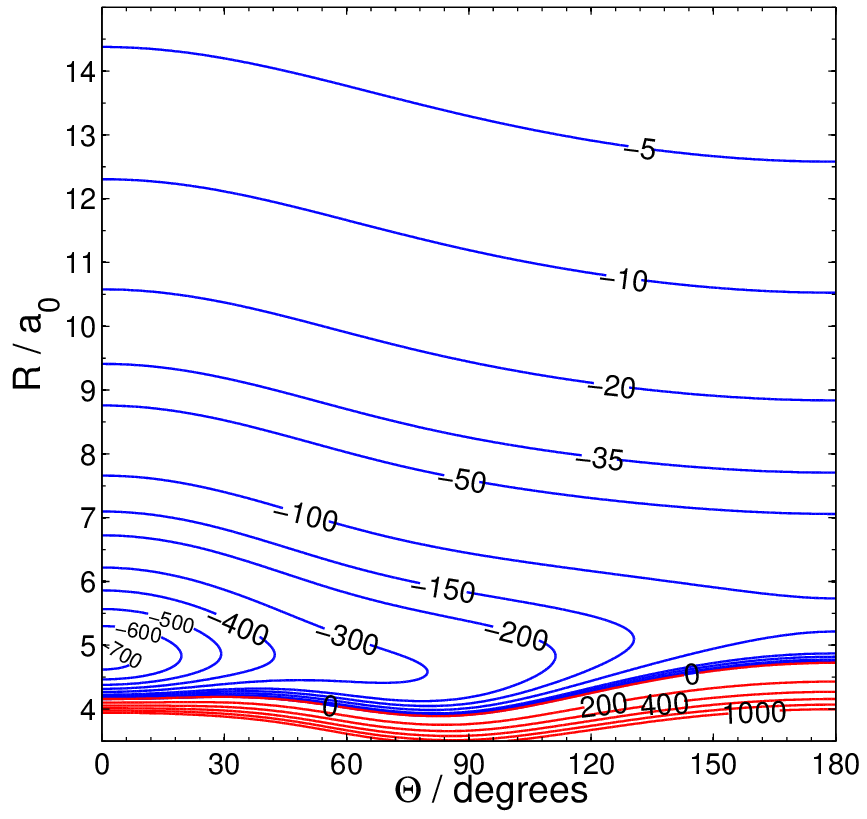} 

\vspace*{-0cm}
\caption{The contour plot of the vibrationally averaged
 $V(R, r, \theta)$ 3-D He-OH$^+$ PES over $\nu=0$
 wave function of the OH$^+$ cation. Contour labels are in cm$^{-1}$}
\label{fig:OHpHe_PES}
\end{figure}

A new PES for  OH$^+$ + He system has been calculated in this work as described in Appendix C.
The average of the three-dimensional PES over the ground
 vibrational state of the OH$^+$ cation  is shown in Figure~\ref{fig:OHpHe_PES}. 
This two-dimensional effective potential is the one used  in subsequent scattering calculations,
described in Appendix D.

\begin{figure}[t]
\begin{center}
\vspace*{-0cm}
\includegraphics[width=8cm,angle=0.]{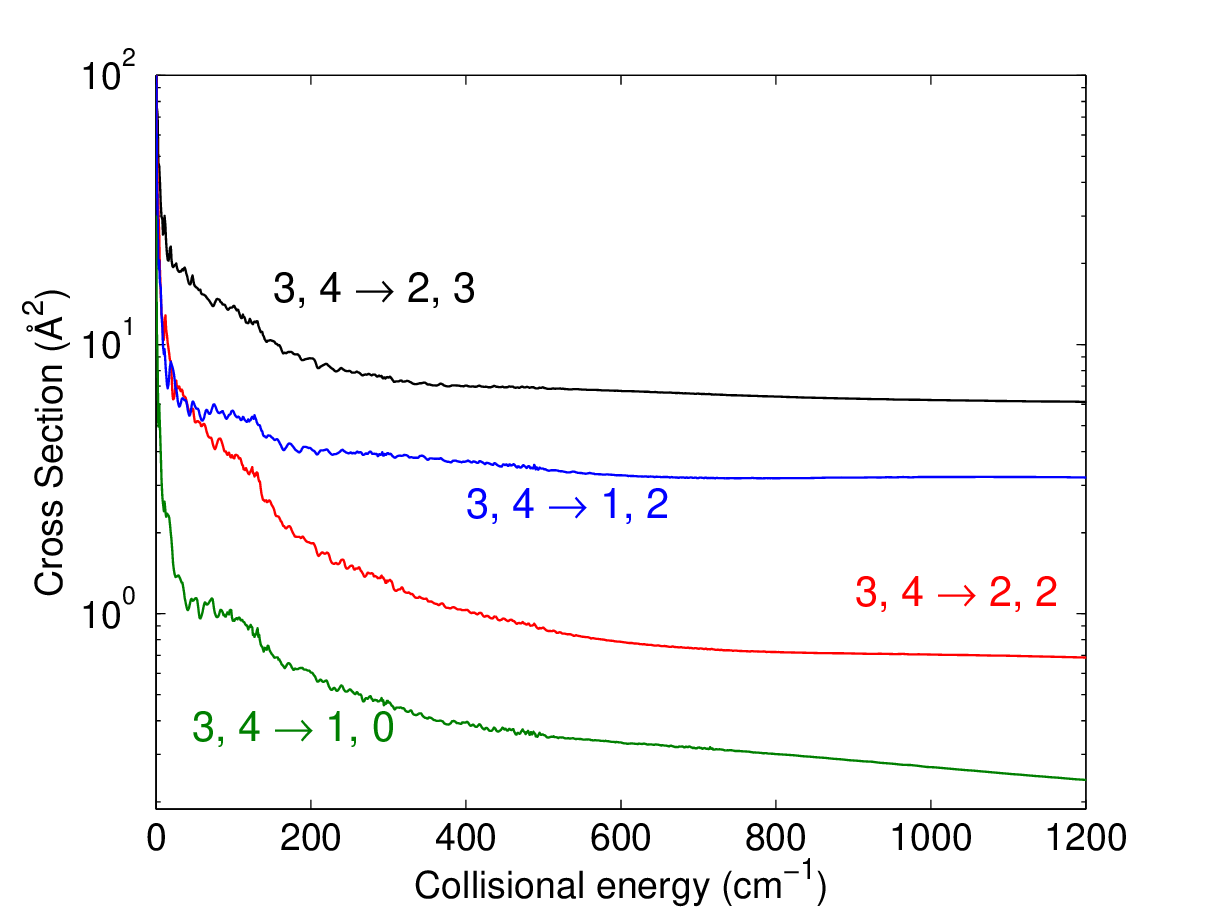}
\vspace*{-0cm}
\caption{Collisional excitation cross sections of OH$^+$ by He out of $N,j= 3,4$ state. 
(a) $\Delta j= \Delta N$ transitions. (b) $\Delta j \ne \Delta N$ transitions.}
\label{fig:xsc}
\end{center}
\end{figure}

We have obtained
the (de-)excitation cross sections for the first 19 fine structure
levels of OH$^+$ by He. Figure \ref{fig:xsc} presents the typical kinetic energy variation of the integral
cross sections for transitions from the fine structure level $(N,
J)=(3, 4)$ of OH$^+$. There are noticeable resonances appearing at low
and intermediate collisional energies. This is related to the presence of an attractive
potential well, which allows for the He
atom to be temporarily trapped there and hence quasi-bound states to
be formed before the complex
dissociates \citep{smith:79,christoffel:90}.

\begin{figure}
\begin{center}
\vspace*{-0cm}
\includegraphics[width=8cm,angle=0.]{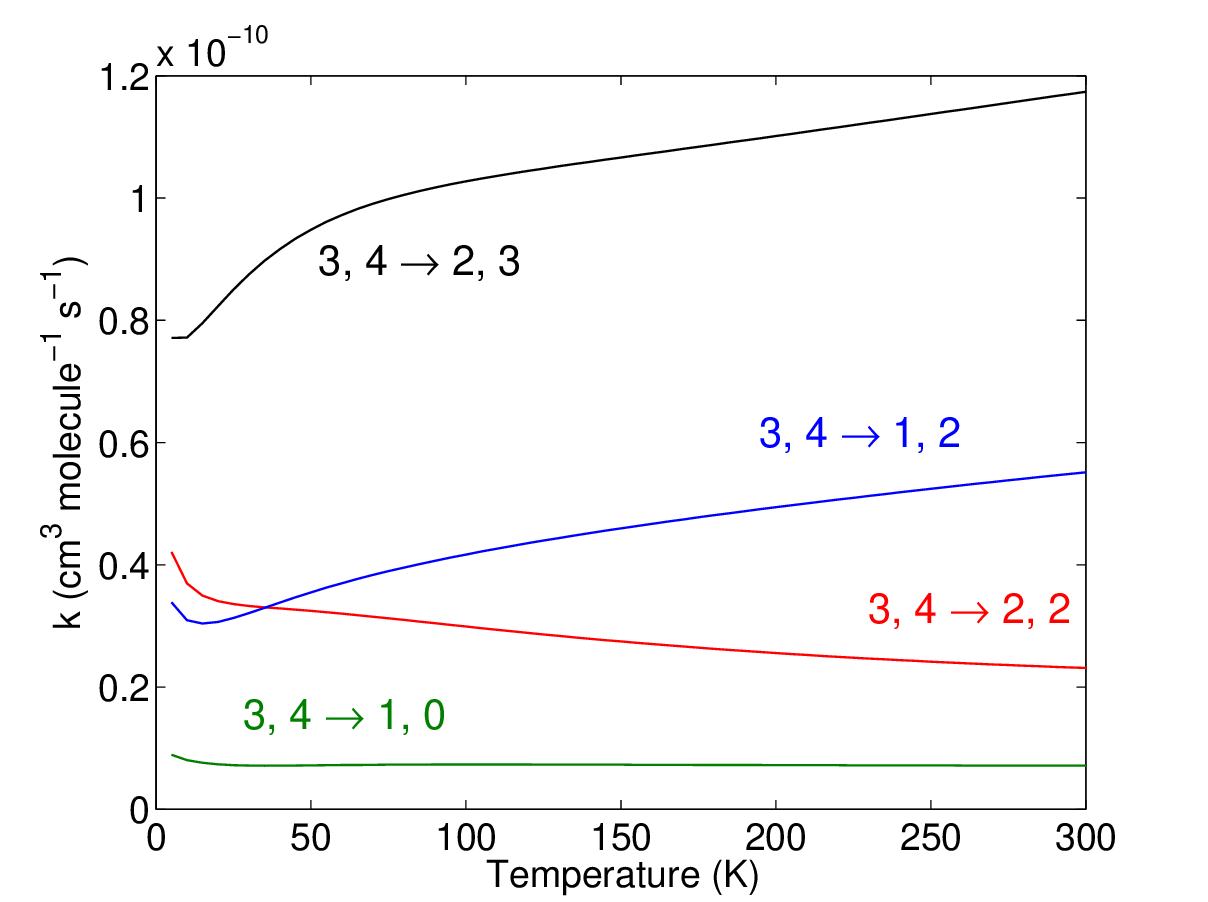}
\vspace*{-0cm}
\caption{Temperature dependence of OH$^+$--He rate constants out of $N,J= 3,4$ state.}
\label{fig:rates}
\end{center}
\end{figure}

By performing a thermal average of the collision energy dependent
cross sections obtained for the first 19 fine-structure OH$^+$ levels, we
obtain rate constants for temperatures up to 300~K. The thermal
dependence of the state-to-state OH$^+$--He rate constants is
illustrated in Fig. \ref{fig:rates} for transitions out of the $N,J= 3,4$
level.

The rate constants shown in Fig. \ref{fig:rates} exhibit interesting features that
have important consequences on the magnitude of fine-structure-resolved rate
constants:

(i) The rate constants decrease with increasing $\Delta N$, 
which is the usual trend for rotational excitation. In addition, 
odd $\Delta N$ transitions are favored over even $\Delta N$ transitions.
 This is a consequence of the strong anisotropy of the PES.

(ii) A strong propensity rule exists for $\Delta J = \Delta N$ transitions.

Such $\Delta J = \Delta N$ propensity rule was
predicted theoretically \citep{alexander:83} and is general for
molecules in the $^{3}\Sigma^{-}$ electronic state. It was also
observed previously for the O$_{2}$(X$^3\Sigma^-$)-He
\citep{lique:10}, SO(X$^3\Sigma^-$)--He \citep{lique:05} or NH(X$^3\Sigma^-$)--He \citep{Tobola:11,Dumouchel:12}
collisions.
  
\begin{figure}[t]
\begin{center}
\vspace*{-0cm}
\includegraphics[width=8.0cm,angle=0.]{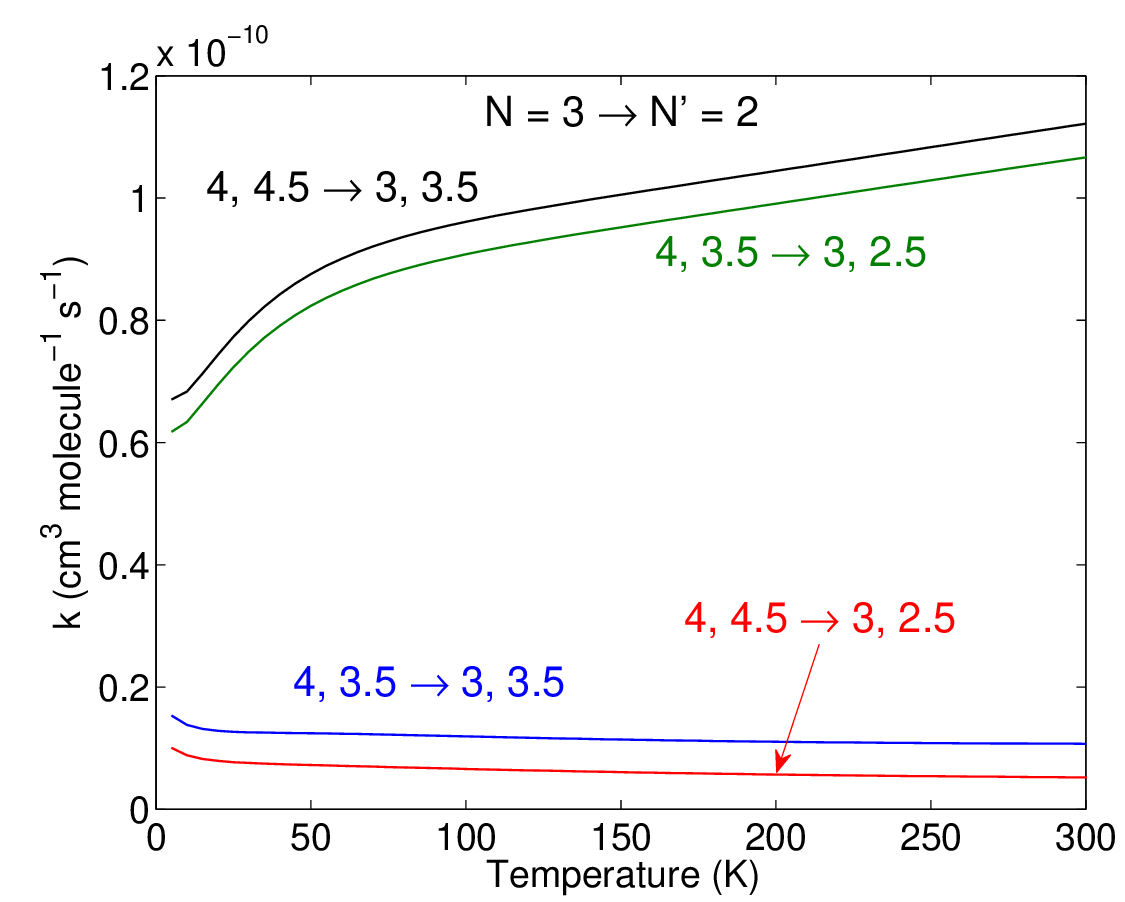}

\vspace*{-0cm}
\includegraphics[width=8.0cm,angle=0.]{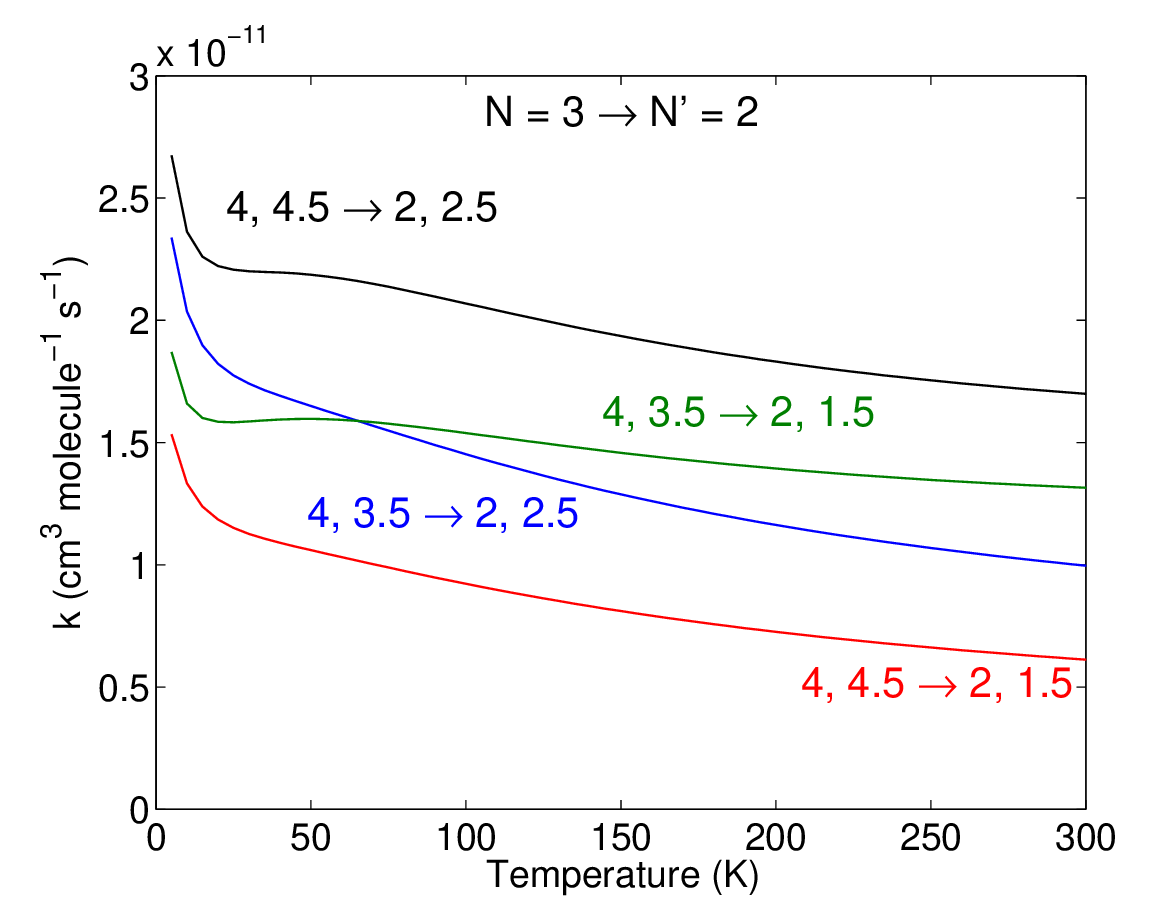}

\vspace*{-0cm}
\caption{Temperature variation of the hyperfine resolved
OH$^+$--He rate constants for $N=3,J=4,F \to
N'=2,J',F'$ transitions. Upper Panel: $\Delta J = \Delta N$ transitions. Lower Panel $\Delta J \ne \Delta N$ transitions.
The numbers correspond to the $J,F \to J',F$ quantum numbers.} \label{fig4}
\end{center}
\end{figure}

Then, we have calculated the hyperfine resolved OH$^+$--He rate coefficients 
using the procedure described in Appendix D. 
The complete set of de-excitation rate coefficients is available
 online from the BASECOL website \citep{Dubernet:13}.
Figure \ref{fig4} presents the temperature variation of the OH$^+$--He rate constants
for selected $N=3,J,F \to N'=2,J',F'$ transitions. 

We have to distinguish $\Delta J = \Delta N$ and $\Delta J \ne \Delta N$
 transitions in order to discuss the hyperfine propensity rules.
For $\Delta J = \Delta N$ transitions, we have a strong propensity rule
 in favor of $\Delta J= \Delta F$ transitions, the propensity rule is 
also more pronounced when the $N$ quantum number increases.
 This trend is the usual trend for open-shell molecules \citep{alexander:85,lique:11CN}.
 For $\Delta J \ne \Delta N$ transitions, it is very difficult to find a clear
 propensity rules as already found for the CN molecule \citep{lique:11CN}.

%%%%%%%%%%%%%%%%%%%%%%%%%%%%%%%%%%%%%%%%%%%%%%%%%%%%%%%%%%%%%%
\section{Applications to astronomical sources}
%%%%%%%%%%%%%%%%%%%%%%%%%%%%%%%%%%%%%%%%%%%%%%%%%%%%%%%%%%%%%%

The computations performed in the previous sections provide an exhaustive dataset of the 
excitation processes of OH$^+$ that may strongly influence the modeling of astronomical 
sources. Such data are indeed critical to understand the physical conditions of regions 
where OH$^+$ is observed in emission, such as hot and dense PDRs (e.g. the Orion Bar, 
\citealt{vanderTak-etal:13}), planetary nebulae \citep{Etxaluze-etal:14,Aleman-etal:14}  and the nuclei of active galaxies \citep{van-der-Werf2010}. 

In these environments, \citet{vanderTak-etal:13} proposed that the formation of OH$^+$ via 
\begin{equation} \label{Eq-reac-OHp}
{\rm O}^+ + {\rm H}_2 \rightarrow {\rm OH}^+ + {\rm H} \quad (\Delta H_0 = - 0.47 \,\, {\rm eV})
\end{equation} 
is sufficiently rapid to compete with (or even dominate) the non-reactive inelastic 
collisions in the excitation of the rotational levels of OH$^+$. Because of lack of 
information, they assumed that the probability of forming OH$^+$ in an excited level 
follows a Boltzmann distribution at a formation temperature $T_{\rm f}=2000$ K, i.e. 
$\sim$ one third of the exothermicity of the above reaction. 

We break here from this approach and treat the chemistry and excitation of OH$^+$ self 
consistently in the framework of the Meudon PDR chemical model \citep{Le-Petit2006,
Le-Bourlot2012} in order to address the following questions. What are the relative 
importances of each process in the excitation of the rotational lines of OH$^+$ ? In 
particular, does the state-to-state rate constants substantially influence the 
emissivities of this species ?

\subsection{Modeling of hot and dense PDRs}

\begin{figure}[!hb]
\begin{center}
\includegraphics[width=7.5cm,angle=-0]{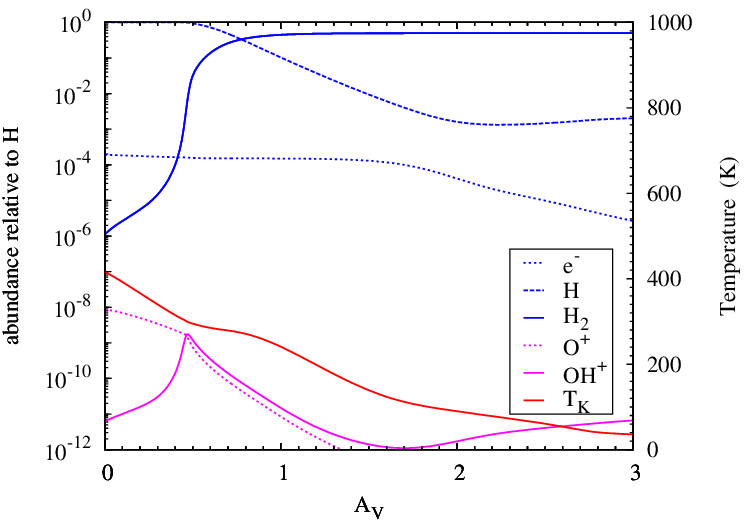}
\caption{Kinetic temperature and abundances relative to total Hydrogen density of H, 
H$_2$, O$^+$, OH$^+$, and $e^-$ computed with the Meudon PDR code as functions of the 
visual extinction from the ionization front across a prototypical hot and dense PDR
($\chi=$10$^4$ and $n$=10$^4$\,cm$^{-3}$).}
\label{Fig-PDR-profil}
\end{center}
\end{figure}

We consider a prototypical hot and dense PDR, i.e. a one-dimensional slab of gas of total
visual extinction $A_{V,{\rm max}}=10$, with a density $n_{\rm H} = 10^4$ cm$^{-3}$, a
cosmic ray ionization rate $\zeta = 3 \times 10^{-16}$ s$^{-1}$ \citep{Indriolo2007,
Indriolo2012} and illuminated from one side by a UV radiation field of $10^4$ that of the 
local ISRF \citep{Mathis1983}. The Meudon PDR code has been run using the standard chemical 
network available online (\url{http://pdr.obspm.fr/PDRcode.html}). The resulting kinetic 
temperature of the gas, its electronic fraction, and the relative abundances of H, H$_2$, 
O$^+$, and OH$^+$ are shown in Fig. \ref{Fig-PDR-profil} as functions of the distance 
from the ionization front. These chemical profiles  indicate that the abundance of OH$^+$ peaks ($n({\rm OH}^+)
\sim 1.8 \times 10^{-5}$ cm$^{-3}$) at $0.3 \leqslant A_V \leqslant 0.7$, i.e. in a region 
where the kinetic temperature $\sim 300$ K, $n(e^-) \sim 1.6$ cm$^{-3}$, $n({\rm He}) 
\sim 10^3$ cm$^{-3}$, $n({\rm O}^+) \sim 1.3 \times 10^{-5}$ cm$^{-3}$, and where most of 
the Hydrogen is in atomic form ($n({\rm H}_2)=10^2$ cm$^{-3}$).

The non reactive inelastic collision rates of OH$^+$ with H, H$_2$, He and $e^-$ have all
been implemented in the Meudon PDR code. For collisions with He we adopt the rates 
computed in the previous sections. For collisions with H and H$_2$, we scale the OH$^+$-He 
collisional rates by using the cross sections calculated for He but using the good
the reduced mass in the thermal average done to calculate the corresponding
rate constants. At last for collisions with 
electrons we adopt the rates of \citet{vanderTak-etal:13} given in their Appendix A and available on
the LAMBDA website \citep{schoier:05} who performed detailed calculations of
the $\Delta N = 1$ transitions. Given the large dipole moment of OH$^+$ we finally assume
$k(\Delta N = 2) = 0.1 \times k(\Delta N = 1)$ and $k(\Delta N > 2) = 0$ \citep{Faure2001}
for higher transitions of the OH$^+$-$e^-$ system.

Concerning the chemical de-excitations, the destruction rates of OH$^+(N,J)$ are supposed 
to be independent from $N,J$ for all the reactions involved in the destruction of OH$^+$. 
Inversely, we assume that the probabilities of forming OH$^+$ in an excited level follow a 
Boltzmann distribution at the kinetic temperature of the gas for all the reactions involved 
in the production of OH$^+$, except for reaction \ref{Eq-reac-OHp} for which we use our 
quantum calculations of the state-to-state rate constants.

\subsection{Intensities of the first rotational lines of OH$^+$}

\begin{figure}[!hb]
\begin{center}
\includegraphics[width=7.5cm,angle=-0]{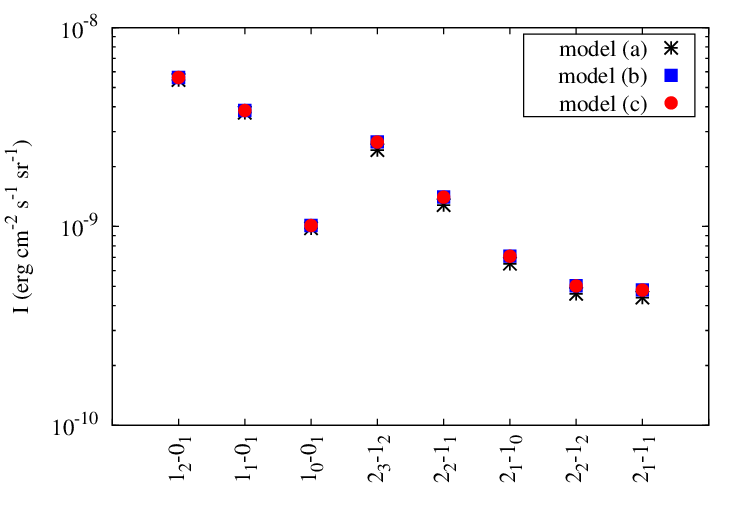}
\caption{Continuum-subtracted intensities of the first rotational lines of OH$^+$ computed 
with the Meudon PDR code in the direction perpendicular to the slab for the models (a), (b), 
and (c) (see main text). Each ${N'}_{J'} - {N''}_{J''}$ line is labeled on the $x$-axis.}
\label{Fig-PDR-lines}
\end{center}
\end{figure}

Following \citet{Zanchet-etal:13}, the Meudon PDR code has been run in three different 
configurations: (a) considering only the excitation by nonreactive collisions, (b) 
including chemical pumping assuming that the probability to form OH$^+$ in an excited 
level via reaction \ref{Eq-reac-OHp} follows a Boltzmann distribution at a formation 
temperature of $2000$ K (as done by \citealt{vanderTak-etal:13}), and (c) adopting the 
branching ratios obtained with our quantum calculations (Table~\ref{tabla-v}). The 
continumm-subtracted intensities in the direction perpendicular to the slab of the 
$N = 1 \rightarrow 0$ and $N = 2 \rightarrow 1$ rotational lines of OH$^+$ are shown 
in Fig. \ref{Fig-PDR-lines}.

The analysis of the main excitation and de-excitation pathways at the peak of OH$^+$ 
abundance shows that the excitation of the $N < 3$ levels is primarily driven by inelastic 
collisions with electrons and atomic Hydrogen. As a result, the intensities of the $N=1-0$ and $N=2-1$ 
transitions predicted by models with density ranging between 10$^{4}$ and 10$^5$ cm$^{-3}$
 increase by less than 10\% when we take the chemical pumping into account. Moreover 
we find no substantial difference between the models computed with
detailed state-to-state chemical rates and those obtained with a Boltzmann distribution 
function. The chemical pumping has a stronger impact on the population of the $N>2$ levels, 
but only in the inner parts of the cloud where the kinetic temperature is lower.

While these results stress the importance of detailed calculations of inelastic collisional 
rates, they do not preclude the existence of interstellar media where the integrated intensities 
of the rotational lines of OH$^+$ may be driven by chemical pumping. Indeed the abundance
of OH$^+$ peaks in a region of the cloud where the abundance of H$_2$ varies over more than
four orders of magnitude (see Fig. \ref{Fig-PDR-profil}). A slightly broader peak of OH$^+$ 
that extends towards the molecular region, as it is the case in media with constant thermal 
pressure rather than constant density \citep{vanderTak-etal:13}, would thus greatly enhance 
the influence of chemical pumping through reaction \ref{Eq-reac-OHp}. To study these effects, 
we will perform a more complete analysis of different astrophysical environments in a 
forthcoming paper. This will be done in the framework of both the Meudon PDR code and the MADEX 
radiative transfer model \citep{Cernicharo2012} in order to also address the impact of  the fluorescence on the excitation of the OH$^+$ rotational lines.

\section{Conclusions}

In this work the state-to-state rate constants for the formation of OH$^+(X^3\Sigma^-)$
products in the reaction O$^+$+H$_2(J=0,1)$ have been obtained using an accurate
quantum wave packet treatment, on the ground electronic state of the system.
In these calculations the electronic spin is not account for, so that it is assumed 
that the rate to form  $F_1(J=N+1)$, $F_2(J=N)$ and $F_3(J=N-1)$ sublevels 
are the same to that of a given final $N$ value obtained here. The results obtained
have been fitted to an analytical form 
in the (0,5000)K temperature interval, and the parameters thus obtained
are listed in the Appendix.

The state-to-state Einstein   coefficient for the $^3\Sigma^-$ - $^3\Sigma^-$,
$^3\Sigma^-$ - $^3\Pi$ and $^3\Pi$ - $^3\Pi$ bands have been calculated and provided 
in the Appendix. For that purpose very accurate
potential energy curves of several electronic states of OH$^+$ have been calculated,
and their corresponding transition dipole moment. 
The empirical  spin-orbit, spin-rotation and spin-spin constants 
have been used \citep{Merer-etal:75,Gruebele-etal:86}.
The rovibrational state on each electronic
state have been calculated and the radial dipole moments have been calculated numerically.
These results are intended to be included in astrophysical PDR model to account for the IR and UV
radiative transfer due to the radiation flux. The radiative lifetimes obtained here, 
of $\approx 2500$ ns,
are in good agreement with the experimental results of \cite{Mohlmann-etal:78}, and 2.5
times  longer than the values reported by 
\cite{Brzozowski-etal:74} which were used by \cite{Almeida-Singh:81} to get semi-empirical Eisntein's
coefficients.

Also collisional OH($X ^3\Sigma^-$) + He inelastic rates have been obtained, including
hyperfine structure and using a new potential energy surface. These are used in the 
astrophysical model used here and also to extrapolate the corresponding rates
for OH($X ^3\Sigma^-$) + H and OH($X ^3\Sigma^-$) + H$_2$. 

All the rates computed in this work have been used in astrophysical models of highly 
illuminated isochoric photodissociation regions. These models show that OH$^+$ is formed
in regions where the kinetic temperature is high and the density of H$_2$ is low. Under
such conditions, we find that chemical pumping does not play a significant role (about
10\% or less) on the excitation of OH$^+$ whose rotational levels are mainly populated through 
inelastic collisions. We propose that chemical pumping may be more efficient if OH$^+$ 
was formed in regions with larger molecular fraction (such as isobaric PDRs) but this 
has yet to be confirmed with additional modeling. Given the importance of inelastic 
collisions, additional computations are now in progress in order to derive more reliable 
estimates of the collisional rates between OH$^+$($^3\Sigma^-$) and both H and H$_2$.

\section{Acknowledgments}

This work has been supported 
by the of Ministerio de Econom{\'\i}a e Innovaci\'on under grants 
CSD2009-00038,   FIS2011-29596-C02 and CTQ2012-37404-C02.
NB acknowledges the Scientific and Technological Council of Turkey for TR-Grid facilities ( TUBITAK; Project No. TBAG-112T827).
ME, JRG and JC thank the Spanish MINECO for funding support from grants AYA2009-07304 and AYA2012-32032.
 OR, AA and NB also acknowledge CSIC 
for a travelling grant I-LINK0775. 
We acknowledge the CNRS national program ``Physique et Chimie du
Milieu Interstellaire'' for supporting this research. F.L. acknowledge support by the Agence Nationale
de la Recherche (ANR-HYDRIDES), contract ANR-12-BS05-0011-01.
J.K. acknowledges the U. S. National Science Foundation (grant
CHE-1213322 to Prof. M. H. Alexander).
The calculations have been perfomed in the parallel facilities at CESGA computing center, 
through ICTS grants, which are acknowledged.

\section{APPENDIX A:
State-to-state reaction rate constants}

The state-to-state reaction rate constants are computed with a time dependent WP method,
 using 
a modified Chebyshev integrator \citep{Huang-etal:94,Mandelshtam-Taylor:95b,Huang-etal:96,%
Kroes-Neuhauser:96,Chen-Guo:96,
Gray-Balint-Kurti:98,Gonzalez-Lezana-etal:05}.
 The WP is represented in reactant Jacobi coordinates in a body-fixed frame, which 
allows to account for the permutation symmetry of H$_2$. At each iteration, a transformation 
to products Jacobi coordinates is performed in order to analyze the final flux on different OH$^+$($v',N$) 
channels, using the method described by \cite{Gomez-Carrasco-Roncero:06}. 
The calculation are performed  using the MAD-WAVE3 
program \citep{Zanchet-etal:09b}. The parameters used in the propagation are  
 listed in Table~\ref{wvp-parameters}.

%%%%%%%%%%%%%%%%%%%%%%%%%%%%%%%%%%%%%%%%%%%%%%%%%%%%%%%%%%%%%%%%%%%%%
 \begin{table}[h]
 \caption{\label{wvp-parameters}
\rm{Parameters used in the wave packet calculations in reactant Jacobi coordinates.
The function used for the absoprtion has the form
$f(X) = \exp\left\lbrack -A_X\left({X-X_{abs}\over b}\right)^n\right\rbrack$ for $X> X_{abs}$ and $f(X) = 1$ elsewhere,
with $X\equiv R$ and $r$, with $n = 4$ and $b=2$. Distances are in \AA\ and
energies in {\rm eV}.}
}
 \begin{center}
 \begin{tabular}{|c|c|}
 \hline 
$r_{min}$, $r_{max}$, $N_r$& 0.2,  30, 256 \\
$r_{abs}$,  $A_r$   & 16,  3 10$^{-6}$   \\
$R_{min}$, $R_{max}$ , $N_R$ & 0.32, 36, 620 \\
$R_{abs}$,  $A_R $    &16 ,  10$^{-6}$ \\
$N_\gamma$ & 160 in $[0,\pi/2]$  \\
$R_0$ , $E_0$ , $\Delta E$   & 13, 0.2,  .1  \\
$R'_\infty$ & 11  \\
$V_{cut}$ & 3.7 \\
$E^{\ell}_{cut}$  & 5 \\
$\Omega_{max}$  & 7 \\
$\Omega'_{max}$  & 25 \\
 \hline
 \end{tabular}
 \end{center}
 \end{table}
%%%%%%%%%%%%%%%%%%%%%%%%%%%%%%%%%%%%%%%%%%%%%%%%%%%%%%%%%%%%%

\begin{figure}[b]
\hspace*{0.5cm}
\includegraphics[scale=0.40,angle=0]{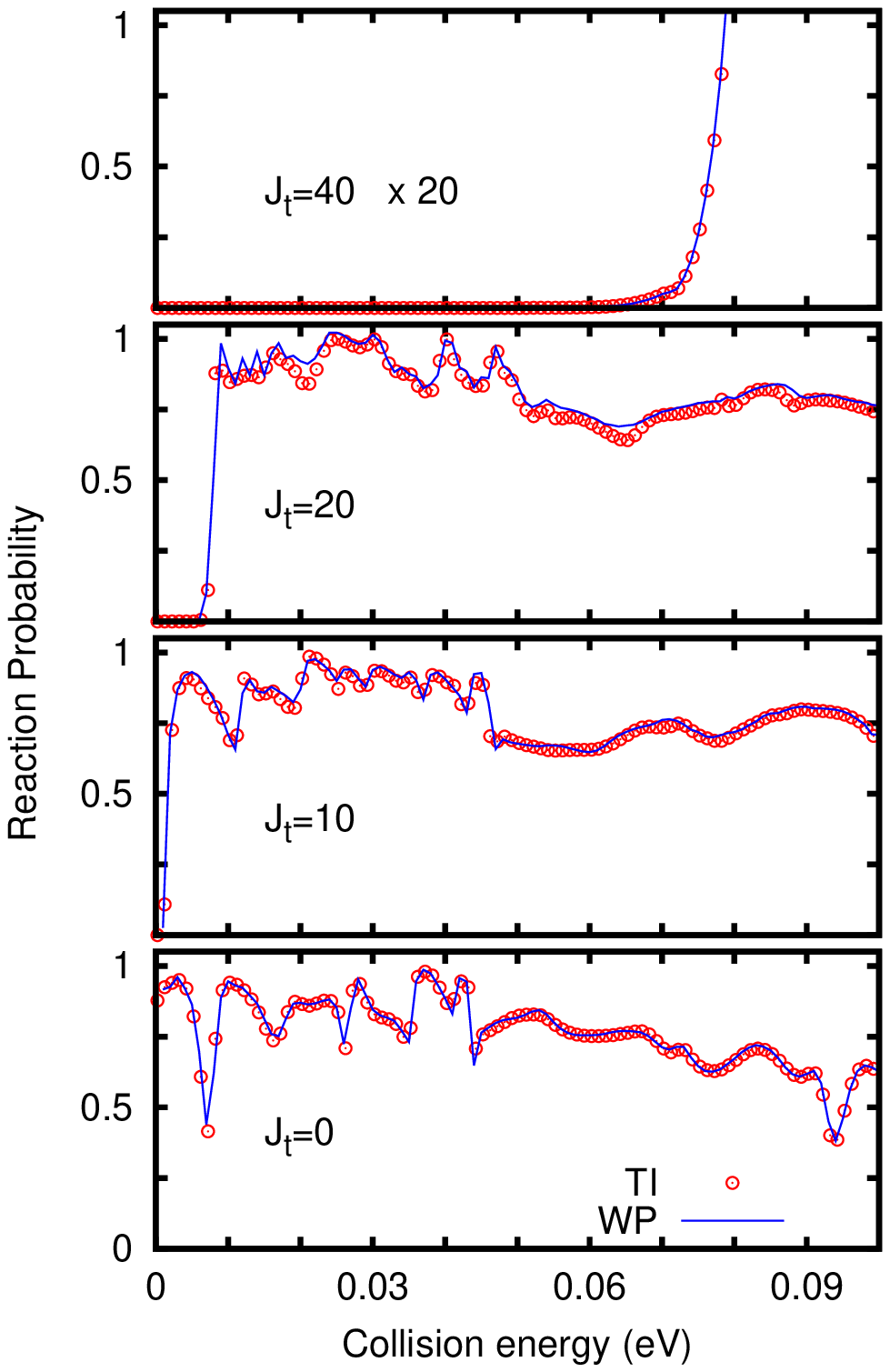}
   \caption{\label{probJ}
Reaction probabilities  for the O$^+$+H$_2(v=0,J=0,J_t)$, for 
different $J_t$ values as a function of collision energy. 
Blue lines are the wave packet results obtained with MADWAVE3 code. 
Open red circles
are the time-independent results obtained with the ABC TI scattering code.
}
\end{figure}

In order to get convergence at  collisional energies of $\approx$ 1 meV, the absorption parameters
have been fitted carefully
 to avoid the reflection of the WP. Also, to get convergence at so low energy,
 a large number  Chebyshev iterations have been performed. This
 number decreases as total angular momentum, $J_t$, increases, because the centrifugal
barrier shifts  the energy threshold towards higher energies and increases the resonances' widths.
More than 100000 iterations were used for $J_t<10$, about 50000 in the interval $10<J_t<20$,
and 30000 or less for $J_t>20$.
The total reaction probabilities in the low  collision energy range 
for some selected $J_t$ values
are compared in Fig.~\ref{probJ} with time-independent (TI) calculations performed using a
coupled channel hyperspherical coordinate method as implemented in the ABC code 
\citep{Skouteris-etal:00}. The comparison shows an excellent agreement down to collision energies lower
than  1 meV.

%%%%%%%%%%%%%%%%%%%%%%%%%%%%%%%%%%%%%%%%%%%%%%%%%%%%%%%%%%%
%%%%%%%%%%%%%%%%%%%%%%%%%%%%%%%%%%%%%%%%%%%%%%%%%%%%%%%%%%%%

The reaction is exothermic and rather fast and, therefore, the Coriolis couplings do not
mix too many helicity states, characterized by the projection of the total angular momentum ${\bf J}_t$
on the z-axis of the body-fixed frame, $\Omega$. As found earlier for this system \citep{Martinez-etal:06b,Xu-etal:12},
 a maximum value of $\Omega_{max} = 7$ is enough to get good convergence. At each iteration
the wave packet is transformed from the reactants to the products body-fixed frame, and in the products
frame a maximum number of $\Omega'_{max}$ = 24 is used. 
In order to control the reactant to product transformation of coordinates,
 the sum of all the individual state-to-state reaction probabilities 
(shown in   Fig.~\ref{probJ}) is compared with that obtained  by the flux 
method \citep{Miller:74,Zhang-Zhang:94a,Neuhauser:94,Goldfield-etal:95}. In all cases, the agreement
is better than 1$\%$.

The reaction probabilities have been calculated for all $J_t$ up to $J_t = 30$. After this value,
only partial waves for $J_t$ in multiple of 5  have been calculated up to $J_t$=80 and 
for all initial helicities $\Omega_0 = 0,...,min(J,J_t)$ and the two parity under inversion 
of spatial coordinates.
For the non-calculated intermediate  $J_t$ values, the reaction probabilities
are obtained using a J-shifting based interpolation, as used before by \citet{Aslan-etal:12,Zanchet-etal:13}.
The convergence of this approach has been tested by comparing the total integral
reaction cross section with that using less number of $J_t$ values, giving an agreement
better than 2$\%$.

The state-to-state rate constants
of the H$_2(v=0, J=0,1)$+ O$^+$ $\rightarrow$  H + OH$^+(v', N)$ reactive collisions 
 have been fitted in the 50-5000 K temperature range to the expression 
\begin{eqnarray}\label{rate-fit}
k_{v',N}(T)= c \, T^b  {\rm exp}(d/T-aT ) \quad 10^{-9} {\rm cm}^3/{\rm s} \nonumber
\end{eqnarray}
where $T$ is in Kelvin.
The parameters $a, b, c,$ and $d$ depend on the initial H$_2(v,J)$ and
 final OH$^+(v',N)$ states and are listed in table~\ref{tabla-v}.

%%%%%%%%%%%%%%%%%%%%%% state-to-state rates table
 \begin{table}[h]
 \caption{\label{tabla-v}
\rm{Parameters used to fit the state-to-state rate constants, according to Eq.~(\ref{rate-fit}),
for O$^+$ + H$_2(v, J) \rightarrow$ H+ OH$^+(v',N)$. 
}}
 \begin{tabular}{|c|c|c|c|c|}
 \hline 
 N & c & b & a $\times$ 1000 & d \\
\hline
\hline
\multicolumn{5} {c} { O$^+$ + H$_2(v=0,J=0) \rightarrow$ H +OH$^+(v'=0, N$) } \\
 \hline 
\hline

   0 &      0.2642E-01 &     0.1125 &     0.2543 &  -128.6513 \\
   1 &      0.7329E-01 &     0.0847 &     0.2407 &  -105.5280 \\
   2 &      0.1144E+00 &     0.0295 &     0.2014 &   -66.0416 \\
   3 &      0.2001E+00 &    -0.0588 &     0.1426 &   -47.4437 \\
   4 &      0.2352E+00 &    -0.0807 &     0.1234 &   -40.9251 \\
   5 &      0.2764E+00 &    -0.1180 &     0.1104 &   -28.9175 \\
   6 &      0.5798E+00 &    -0.2418 &     0.0659 &   -27.7198 \\
   7 &      0.5885E+00 &    -0.2520 &     0.0592 &   -24.1934 \\
   8 &      0.4493E+00 &    -0.2222 &     0.0668 &   -23.3146 \\
   9 &      0.4858E+00 &    -0.2387 &     0.0708 &   -29.2944 \\
  10 &      0.3966E+00 &    -0.2170 &     0.0711 &   -28.6389 \\
  11 &      0.2131E+00 &    -0.1421 &     0.0807 &   -33.7970 \\
  12 &      0.1478E+00 &    -0.1126 &     0.0799 &   -37.3616 \\
  13 &      0.5621E-01 &    -0.0247 &     0.0748 &   -22.3201 \\
  14 &      0.3229E-01 &     0.0140 &     0.0685 &   -32.7974 \\
  15 &      0.2872E-02 &     0.2924 &     0.1120 &   -67.0153 \\
  16 &      0.4806E-06 &     1.4175 &     0.3711 &    -0.0028 \\
  17 &      0.1200E-08 &     2.1348 &     0.4843 &    -0.0027 \\
  18 &      0.5969E-11 &     2.7550 &     0.5812 &    -0.0028 \\
  19 &      0.7767E-13 &     3.2592 &     0.6782 &    -0.0032 \\
  20 &      0.1965E-15 &     3.6601 &     0.7370 &    -0.0032 \\
  21 &      0.3068E-16 &     4.1274 &     0.8156 &    -0.0035 \\
  22 &      0.1796E-17 &     4.3997 &     0.8392 &    -0.0034 \\
\hline
\hline
\multicolumn{5} {c} { O$^+$ +H$_2(v=0,J=0) \rightarrow$ H +OH$^+(v'=1, N$) } \\
\hline 
\hline

   0 &      0.7654E-01 &    -0.1131 &     0.1579 &   -93.1416 \\
   1 &      0.6630E-01 &     0.0100 &     0.1786 &   -52.6948 \\
   2 &      0.6690E-01 &    -0.0015 &     0.1414 &   -33.0402 \\
   3 &      0.2440E+00 &    -0.1824 &     0.0821 &   -49.1025 \\
   4 &      0.3277E+00 &    -0.2307 &     0.0562 &   -36.2931 \\
   5 &      0.3474E+00 &    -0.2639 &     0.0220 &   -24.5429 \\
   6 &      0.5062E+00 &    -0.3307 &     0.0050 &   -31.8403 \\
   7 &      0.1592E+00 &    -0.1843 &     0.0516 &  -143.9353 \\
   8 &      0.8583E-01 &    -0.1036 &     0.0914 &  -518.6398 \\
   9 &      0.1506E+00 &    -0.2041 &     0.0690 &  -908.0244 \\
  10 &      0.6036E-06 &     1.4279 &     0.4508 &    -0.0039 \\
  11 &      0.1361E-07 &     1.8738 &     0.4999 &    -0.0028 \\
  12 &      0.5250E-09 &     2.2663 &     0.5603 &    -0.0029 \\
  13 &      0.2768E-10 &     2.6122 &     0.6105 &    -0.0030 \\
  14 &      0.1386E-11 &     2.9556 &     0.6522 &    -0.0030 \\
  15 &      0.2221E-12 &     3.1393 &     0.6501 &    -0.0029 \\
  16 &      5.0251E-18 &     4.5027 &     0.8880 &    -0.0001 \\
  17 &      3.2730E-18 &     4.5027 &     0.8880 &    -0.0001 \\
  18 &      0.2130E-17 &     4.5027 &     0.8801 &    -0.0036 \\
  19 &      0.1380E-18 &     4.7881 &     0.9153 &    -0.0037 \\
  20 &      0.5397E-18 &     4.4867 &     0.7874 &    -0.0033 \\
\hline
\end{tabular}
\end{table}
\begin{table}[h]
 \begin{tabular}{|c|c|c|c|c|}

\hline
\hline
\multicolumn{5} {c} { O$^+$ + H$_2(v=0,J=1) \rightarrow$ H +OH$^+(v'=0, N$) } \\
 \hline 
\hline
   0 &      0.4216E-02 &     0.2224 &     0.2236 &   -23.2949 \\
   1 &      0.1089E-01 &     0.2557 &     0.2377 &   -19.5720 \\
   2 &      0.2190E-01 &     0.2270 &     0.2318 &   -20.3101 \\
   3 &      0.6200E-01 &     0.1099 &     0.1939 &   -25.9693 \\
   4 &      0.1882E+00 &    -0.0349 &     0.1377 &   -34.3086 \\
   5 &      0.3235E+00 &    -0.1204 &     0.0931 &   -35.2714 \\
   6 &      0.3053E+00 &    -0.1404 &     0.0726 &   -24.7791 \\
   7 &      0.4544E+00 &    -0.2133 &     0.0572 &   -24.0813 \\
   8 &      0.3988E+00 &    -0.2134 &     0.0564 &   -24.4975 \\
   9 &      0.2089E+00 &    -0.1394 &     0.0831 &   -19.5044 \\
  10 &      0.2260E+00 &    -0.1534 &     0.0885 &   -28.3408 \\
  11 &      0.1936E+00 &    -0.1428 &     0.0796 &   -28.4596 \\
  12 &      0.6348E-01 &    -0.0158 &     0.0944 &   -17.7454 \\
  13 &      0.2635E-01 &     0.0671 &     0.1025 &   -18.2581 \\
  14 &      0.1827E-01 &     0.0649 &     0.0788 &   -29.9789 \\
  15 &      0.1409E-02 &     0.3549 &     0.1156 &   -20.4638 \\
  16 &      0.8814E-06 &     1.3023 &     0.3317 &    -0.0026 \\
  17 &      0.5001E-08 &     1.9172 &     0.4424 &    -0.0025 \\
  18 &      0.1706E-10 &     2.6009 &     0.5720 &    -0.0030 \\
  19 &      0.1556E-12 &     3.1589 &     0.6664 &    -0.0032 \\
  20 &      0.3814E-06 &     0.9388 &    -0.0935 &    -0.0009 \\
  21 &      0.9051E-16 &     3.9766 &     0.8019 &    -0.0034 \\
  22 &      0.8370E-17 &     4.1892 &     0.8050 &    -0.0033 \\
\hline
\hline
\multicolumn{5} {c} { O$^+$ +H$_2(v=0,J=1) \rightarrow$ H +OH$^+(v'=1, N$)} \\
 \hline 
\hline
   0 &      0.3753E-01 &    -0.1096 &     0.1175 &  -126.3446 \\
   1 &      0.1323E+00 &    -0.1232 &     0.1292 &  -153.8032 \\
   2 &      0.4300E-01 &     0.0738 &     0.1825 &   -59.5045 \\
   3 &      0.6226E-01 &     0.0309 &     0.1558 &   -32.2931 \\
   4 &      0.4331E+00 &    -0.2623 &     0.0362 &   -42.6035 \\
   5 &      0.3119E+00 &    -0.2377 &     0.0372 &   -33.0488 \\
   6 &      0.3304E+00 &    -0.2614 &     0.0324 &   -20.8766 \\
   7 &      0.3757E+00 &    -0.3033 &     0.0213 &   -39.8215 \\
   8 &      0.2547E+00 &    -0.2644 &     0.0405 &  -352.3149 \\
   9 &      0.2390E-02 &     0.3403 &     0.1918 &  -373.4869 \\
  10 &      0.7570E-05 &     1.0464 &     0.2998 &    -0.0017 \\
  11 &      0.8401E-07 &     1.6516 &     0.4783 &    -0.0029 \\
  12 &      0.1604E-08 &     2.1177 &     0.5401 &    -0.0029 \\
  13 &      0.9677E-10 &     2.4569 &     0.5985 &    -0.0029 \\
  14 &      0.4207E-11 &     2.8035 &     0.6315 &    -0.0029 \\
  15 &      0.3221E-18 &     4.3612 &     1.0000 &    -0.0010 \\
  16 &      0.2129E-18 &     4.3612 &     1.0002 &    -0.0010 \\
  17 &      0.1392E-17 &     4.3600 &     1.0001 &    -0.0010 \\
  18 &      0.5176E-17 &     4.3612 &     0.8616 &    -0.0036 \\
  19 &      0.3496E-18 &     4.6391 &     0.8928 &    -0.0036 \\
  20 &      0.1317E-15 &     3.6978 &     0.5835 &    -0.0028 \\
\hline
 \end{tabular}
 \end{table}

\newpage

\newpage

\section{APPENDIX B: OH$^+$ Einstein coefficientes}

In order to incorporate electronic transitions in the radiative models, we have calculated 
seven potential energy curves of the OH$^{+}$ cation, correlating with the O($^3P$)+H$^{+}$, 
O$^{+}$($^4S$)+H($^2S$) and O($^1D$)+H$^{+}$ dissociation channels, using the 
MOLPRO package \citep{MOLPRO}  for {\it ab initio} electronic calculations. The calculations initially 
consisted of a full valence state-averaged complete active space procedure (SA-CASSCF) 
including all the molecular orbitals arising from the valence atomic orbitals (8 electrons 
in 5 orbitals). The C$_{2v}$ point group of symmetry has been used. The state-averaged 
electronic wavefunction included all 
the states correlating with the above-mentioned asymptotes, 
two $^3\Sigma^{-}$, one $^3\Pi$, one $^1\Sigma^{+}$, one $^1\Delta$, 
one $^1\Pi$ and one $^5\Sigma^{-}$ electronic states. 
These number of states ensures the correct
 degeneracy of the states at the dissociation limits. These wavefunctions 
were used as reference for a subsequent internally contracted multireference
 configuration interaction (icMRCI) calculation, where all single and
 double excitations were included. Finally,
 the Davidson correction (+Q) \citep{Davidson:75} was applied to the final
 energies in order to approximately
 account for the contribution of higher excitations. Calculations have
 been performed with three correlation-consistent polarized basis set of
 Dunning, denoted aug-cc-pVnZ ($n = $ Q, 5 and 6), and the extrapolation to 
complete basis set (CBS, $n= \infty$) has been also obtained using the
 following extrapolation formula \citep{Woon-Duninng:94}: 
\begin{eqnarray}
\nonumber
E(n) = E_{CBS} + B e^{-(n-1)} + C e^{-(n-1)^2}.
\end{eqnarray}
The CBS electronic energy curves
  are displayed in the top panel of Fig.~\ref{OHp_electronic_states}.

In order to calculate the ro-vibrational
states of OH$^+$, we use the effective Hamiltonian \citep{Lefebvre-Brion-Field:86}
\begin{eqnarray}\label{effective-hamiltonian}
H&=&-{\hbar^2\over 2\mu}{1\over R^2}{\partial\over\partial R}R^2{\partial\over\partial R}
+{{\bf R}^2\over 2\mu R^2} \nonumber\\
&+&{2\over 3} \lambda \left(3{\bf S}_z^2 -{\bf S}^2\right)
+\gamma {\bf R}\cdot {\bf S}
+V,
\end{eqnarray}
where the third and forth  terms are the spin-spin and the
 spin-rotation terms. 
In this equation $V$ is the {\it ab initio} potential
energy curves  calculated for the $X^3\Sigma^-$ and $A^3\Pi$ states.
The spin-orbit is included as an empirical parameter taken from
 \cite{Merer-etal:75}.
The X$^3\Sigma^-$ state is represented in the  Hund's case (b),
with the parameters $\gamma$= -0.15126 cm$^{-1}$ and $\lambda=2.1429$ cm$^{-1}$
as determined in measurements of the rotational spectra \citep{Gruebele-etal:86}.
The A$^3\Pi$ states are represented in the  Hund's case (a), with  $\gamma =  0.01730$ cm$^{-1}$
 taken from  \cite{Merer-etal:75}.

The total wavefunction of OH$^+$ is factorized as
\begin{eqnarray}\label{total-diatomic-functions}
\Psi^{JM;S\Lambda}_{\alpha,v}= {\Phi_{v}^{JS\Lambda; \alpha}(R) \over R }
\left\vert J M S \Lambda; \alpha\right\rangle,
\end{eqnarray}
where the quantum number  $\alpha=N$ for  Hund's case (b) to describe
 $X^3\Sigma^-$, while $\alpha=\Sigma$ or $\Omega=\Lambda+\Sigma$ 
for Hund's case (a) for $A^3\Pi$.

The radial functions in Eq.~(\ref{total-diatomic-functions})
are the numerical solutions of the one-dimensional Schr\"odinger equation
\begin{eqnarray}
&&\left\lbrace
-{\hbar^2\over 2\mu}{d^2\over d R^2}
+\hbar^2{ C\over 2\mu R^2}
 + V(R) -E^{JS\Lambda\alpha}_{v}\right\rbrace \varphi^{JS\Lambda;\alpha}_{v}(R)
=0.\nonumber\\
\end{eqnarray}
with 
\begin{eqnarray}
  C=\left\lbrace
     \begin{array}{ccc}
      N(N+1)-\Lambda^2 &{\rm for} & ^3\Sigma^-\quad {\rm case\, (b)}\\
     J(J+1) -\Omega^2 + S(S+1)- \Sigma^2 &{\rm for} & ^3\Pi\quad {\rm case\, (a)}
     \end{array}
  \right.
\nonumber
\end{eqnarray}
The rovibrational states have been obtained numerically, in a grid
of 12000 points in the $(0.4, 14)$ a.u. interval. 
Vibrational levels up to $v=20$ and $v=12$ have been considered
for $X$ and $A$ states, respectively. The higher rotational level
considered are  $N=35$, for $^3\Sigma^-$, and $J=35$ for $^3\Pi_\Omega$.

In Hund's case (b), the angular functions in Eq.~(\ref{total-diatomic-functions}) are defined as
\begin{eqnarray}\label{Hund-b-case}
\left\vert J M S \Lambda; N\right\rangle
  &=& \sum_{M_S,M_N} (-1)^{S-N+M}\sqrt{2J+1} 
\left(\begin{array}{ccc}
  S & N & J\\
  M_S & M_N & M
 \end{array}\right) \nonumber\\
&\times&
\quad\quad\quad\vert SM_S\rangle \left\vert N M_N \Lambda \right\rangle 
\end{eqnarray}
where $M$, $M_S$ and $M_N$  are the projections of ${\bf J}$, ${\bf S}$ and ${\bf N}$ 
 angular momenta, respectively,
 on the z-axis of the laboratory frame. $\vert SM_S\rangle $ are the spin functions and
\begin{eqnarray}
\left\vert N M_N \Lambda \right\rangle =\sqrt{{2N+1\over 4\pi}} D^{N*}_{M_N\Lambda}(\phi,\theta,0) \left\vert \Lambda \right\rangle
\end{eqnarray}
where $D^{N*}_{M_N\Lambda}$ are  Wigner rotation matrices \citep{Zare-book}
and $\Lambda$ is the projection of the electronic orbital angular momentum 
on the internuclear axis, used to label the electronic state $\left\vert \Lambda \right\rangle$.

In Hund's case (a) the angular functions are
\begin{eqnarray}\label{Hund-a-case}
\left\vert J M S \Lambda; \Sigma\right\rangle = 
\sqrt{2J+1\over 4\pi} D^{J*}_{M\Omega}(\phi,\theta,0) \left\vert \Lambda\right\rangle
\left\vert S,\Sigma\right\rangle ,
\end{eqnarray}
where $\Sigma$ is the projections of the electronic spin  on the internuclear axis,
and $\Omega=\Lambda+\Sigma$.

The transformation between the two angular basis sets is given by
\begin{eqnarray}\label{transformation-b2a-hund}
\left\vert J M S \Lambda; N\right\rangle
&=& \sum_{\Sigma} (-1)^{S-N+\Omega}  \sqrt{2N+1}
\left(\begin{array}{ccc}
        S & N & J\\
       \Sigma & \Lambda & - \Omega
\end{array}\right) \nonumber\\
 && \quad\quad\quad \left\vert J M S \Lambda; \Sigma\right\rangle
\end{eqnarray}
 
The total energy, considering spin-spin and spin-rotation terms in the Hamiltonian,
for the $^3\Sigma^-$ states in the b Hund's case 
are given by \citep{Herzberg-diatomics}

\begin{eqnarray}
F_1 &=& E^{JS\Lambda N}_v -{2\over 3} \lambda {N\over 2N+3}  + \gamma N  \quad{\rm with } \quad J=N+1
\nonumber\\
F_2 &=&  E^{JS\Lambda N}_v  +{2\over 3} \lambda  -\gamma  \quad{\rm with } \quad J=N
\\
F_3 &=&  E^{JS\Lambda N}_v   -{2\over 3} \lambda {N+1\over 2N-1}  -\gamma (N+1) \quad{\rm with } \quad J=N-1
\nonumber
\end{eqnarray}
and for the $^3\Pi$ states in the Hund's case a are given by
\begin{eqnarray}
F_3 &=& E^{JS\Lambda \Sigma=-1}_v -  \gamma +\Delta T_0 \quad{\rm with } \quad \Omega=0
\nonumber\\
F_2 &=&  E^{JS\Lambda \Sigma=0}_v  -2\gamma +\Delta T_1    \quad{\rm with } \quad \Omega=1
\\
F_1 &=&  E^{JS\Lambda \Sigma=1}_v  -\gamma   +\Delta T_2 \quad{\rm with } \quad  \Omega=2.
\nonumber
\end{eqnarray}
with $\Delta T_\Omega$ = 0, 83 and 167 cm$^{-1}$, for $\Omega=0,1$ and 2, respectively, associated
to the empirical spin-orbit splittings \citep{Merer-etal:75}.
These three states are doubly degenerate, $g=2$, and in the case of $\Omega=0$ we do consider
the 0$^+$ and 0$^-$ states as degenerate.

The Einstein coefficients are given by
\begin{eqnarray}\label{Einstein-coefficients}
A_{Jv;J'v'}^{\Lambda S\Sigma; \Lambda' S'\Sigma'}
&=& {1 \over 3\pi\epsilon_0 \hbar^4} \left({h\nu \over c}\right)^3 
  \delta_{SS'}
 \left\vert M_{v;v'}^{J\Lambda S \alpha; J' \Lambda' S'\alpha'} \right\vert^2 \nonumber\\
&\times& {S^{J\Lambda S;J' \Lambda' S' }_{\alpha\alpha'}\over 2J+1},
\end{eqnarray}
where the radial integrals are given by 
\begin{eqnarray}
M_{v;v'}^{J\Lambda S \alpha; J' \Lambda' S'\alpha'}
=\int dR\, \Phi^{J\Lambda S\alpha*}_{v}(R)
\left\langle \Lambda\vert d_q \vert \Lambda'\right\rangle
 \Phi^{J'\Lambda' S'\alpha'}_{v'}(R),\nonumber
\end{eqnarray}
and are perfomed numerically using 
$\left\langle \Lambda\vert d_q \vert \Lambda'\right\rangle$ values
calculated with the MOLPRO package.
 The $M_{v;v'}^{J\Lambda S \alpha; J' \Lambda' S'\alpha'}$
matrix elements obtained in this work are also listed for  different
electronic transitions,  $X^3\Sigma^- \leftarrow X^3\Sigma^-$,
  $A^3\Pi_\Omega \leftarrow A^3\Pi_\Omega$ and   $A^3\Pi_\Omega \leftarrow X^3\Sigma^-$,
in the tables listed below.

The H\"onl-London factors
 in Eq.~(\ref{Einstein-coefficients}) depend
on the Hund's case used, and becomes:
\begin{eqnarray}
S^{J\Lambda S;J' \Lambda' S' }_{\alpha\alpha'}&=&
       {(2J+1)(2J'+1)(2N+1)(2N'+1)\over 3}\nonumber\\
    &\times&     \left(\begin{array}{ccc}
              N& 1 & N'\\
              -\Lambda & q & \Lambda'
         \end{array}\right)^2
         \left\lbrace\begin{array}{ccc} 
             J& N & S\\
             N' & J' & 1
          \end{array}\right\rbrace^2
\end{eqnarray}
for $ ^3\Sigma (b)\leftarrow^3\Sigma (b)$ transitions,

\begin{eqnarray}
S^{J\Lambda S;J' \Lambda' S' }_{\alpha\alpha'}=
       {(2J+1)(2J'+1)\over 3}
           \left(\begin{array}{ccc} 
                 J & 1 & J'\\
                 \Omega & q & -\Omega'
            \end{array}\right)^2
\end{eqnarray}
for $ ^3\Pi (a) \leftarrow^3\Pi (a)$, and
\begin{eqnarray}
S^{J\Lambda S;J' \Lambda' S' }_{\alpha\alpha'}&=&
       {(2J+1)(2J'+1)(2N+1)\over 3} \\
&\times&         \left(\begin{array}{ccc}
              J& 1 & J'\\
              \Lambda+\Sigma & q & -\Lambda'-\Sigma
         \end{array}\right)^2
         \left(\begin{array}{ccc}
              S& N & J\\
              \Sigma & \Lambda & -\Lambda-\Sigma
         \end{array}\right)^2 \nonumber
\end{eqnarray}
for mixed
      $ ^3\Pi (a) \leftarrow^3\Sigma (b)$ transitions.

In the tables below  we report 
total energies and the electric dipole moments, 
$\left\vert M_{v;v'}^{J\Lambda S \alpha; J' \Lambda' S'\alpha'} \right\vert^2$
required to calculate the Einstein coefficients in Eq.~(\ref{Einstein-coefficients})
 for different
electronic transitions:  $X^3\Sigma^- \leftarrow X^3\Sigma^-$,
  $A^3\Pi_\Omega \leftarrow A^3\Pi_\Omega$ and   $A^3\Pi_\Omega \leftarrow X^3\Sigma^-$.

In order to reduce the length of the table, only those values 
for $\left\vert M_{v;v'}^{J\Lambda S \alpha; J' \Lambda' S'\alpha'} \right\vert^2 > 0.1$
 (10$^{-3}$   for $\Sigma-\Pi$ transitions)
and since the dependence of  $\left\vert M_{v;v'}^{J\Lambda S \alpha; J' \Lambda' S'\alpha'} \right\vert^2 $
on the rotational quantum numbers ${J \alpha; J' \alpha'}$ is weak, we shall only list
the matrix elements for the lower rotational transition which are larger than $10^{-3}$ for each
electronic transition considered here. The H\"onl-London factors, containing the main
dependence of the Einstein coefficients on the rotational transitions,  can be easily evaluated 
using the expressions of this appendix.

\renewcommand{\arraystretch}{2.5}

\begin{center}
\begin{longtable}{|cc|cc|c|}
\caption{\label{sigma-sigma-transitions} 
\rm{$^3\Sigma^- -^3\Sigma^-$ transition
electric dipole matrix elements (in a.u.)
for $N_X$=0, $J_X$=1,  $N'_X$=1 ,  $J'_X$=0}}\\

\hline
\multicolumn{1}{|c}{} &\multicolumn{1}{c|}{} &\multicolumn{1}{|c}{} &\multicolumn{1}{c|}{} &\multicolumn{1}{c|}{} \\[2pt]
\multicolumn{1}{|c} { $v_X$}  & \multicolumn{1}{c|} {$E_{vNJ}$(cm$^{-1}$)}      & 
 \multicolumn{1}{|c}{ $v'_X$} & \multicolumn{1}{c|} {$E_{v'N'J'}$(cm$^{-1}$)} &
 \multicolumn{1}{c|}{ $\vert M\vert^2$ (a.u.)} \\[2pt]
\multicolumn{1}{|c}{} &\multicolumn{1}{c|}{} &\multicolumn{1}{|c}{} &\multicolumn{1}{c|}{} &\multicolumn{1}{c|}{} \\[2pt]
\hline
\multicolumn{1}{|c}{} &\multicolumn{1}{c|}{} &\multicolumn{1}{|c}{} &\multicolumn{1}{c|}{} &\multicolumn{1}{c|}{} \\[2pt]
\endfirsthead

\hline
\multicolumn{5}{|c|} {  }\\[1pt]
 \multicolumn{5}{|c|} {\rm{ $^3\Sigma^- -^3\Sigma^-$ transitions (continuation)} }\\[2pt]
\multicolumn{5}{|c|} {  }\\[1pt]
\hline
\multicolumn{1}{|c}{} &\multicolumn{1}{c|}{} &\multicolumn{1}{|c}{} &\multicolumn{1}{c|}{} &\multicolumn{1}{c|}{} \\[2pt]

\endhead
   0 &         1540.906 &    0 &         1571.238 &            0.832 \\[0.001mm]
   0 &         1540.906 &    1 &         4528.533 &            0.005 \\[0.001mm]
   1 &         4499.658 &    0 &         1571.238 &            0.005 \\[0.001mm]
   1 &         4499.658 &    1 &         4528.533 &            0.911 \\[0.001mm]
   1 &         4499.658 &    2 &         7329.110 &            0.010 \\[0.001mm]
   2 &         7301.650 &    1 &         4528.533 &            0.011 \\[0.001mm]
   2 &         7301.650 &    2 &         7329.110 &            0.997 \\[0.001mm]
   2 &         7301.650 &    3 &         9979.210 &            0.016 \\[0.001mm]
   3 &         9953.124 &    2 &         7329.110 &            0.018 \\[0.001mm]
   3 &         9953.124 &    3 &         9979.210 &            1.091 \\[0.001mm]
   3 &         9953.124 &    4 &        12484.909 &            0.022 \\[0.001mm]
   4 &        12460.153 &    3 &         9979.210 &            0.025 \\[0.001mm]
   4 &        12460.153 &    4 &        12484.909 &            1.193 \\[0.001mm]
   4 &        12460.153 &    5 &        14852.246 &            0.029 \\[0.001mm]
   5 &        14828.776 &    4 &        12484.909 &            0.032 \\[0.001mm]
   5 &        14828.776 &    5 &        14852.246 &            1.303 \\[0.001mm]
   5 &        14828.776 &    6 &        17087.175 &            0.036 \\[0.001mm]
   5 &        14828.776 &    7 &        19195.549 &            0.001 \\[0.001mm]
   6 &        17064.943 &    5 &        14852.246 &            0.040 \\[0.001mm]
   6 &        17064.943 &    6 &        17087.175 &            1.422 \\[0.001mm]
   6 &        17064.943 &    7 &        19195.549 &            0.043 \\[0.001mm]
   6 &        17064.943 &    8 &        21183.090 &            0.002 \\[0.001mm]
   7 &        19174.511 &    5 &        14852.246 &            0.001 \\[0.001mm]
   7 &        19174.511 &    6 &        17087.175 &            0.048 \\[0.001mm]
   7 &        19174.511 &    7 &        19195.549 &            1.549 \\[0.001mm]
   7 &        19174.511 &    8 &        21183.090 &            0.051 \\[0.001mm]
   7 &        19174.511 &    9 &        23055.314 &            0.003 \\[0.001mm]
   8 &        21163.200 &    6 &        17087.175 &            0.002 \\[0.001mm]
   8 &        21163.200 &    7 &        19195.549 &            0.056 \\[0.001mm]
   8 &        21163.200 &    8 &        21183.090 &            1.684 \\[0.001mm]
   8 &        21163.200 &    9 &        23055.314 &            0.058 \\[0.001mm]
   8 &        21163.200 &   10 &        24817.468 &            0.004 \\[0.001mm]
   9 &        23036.530 &    7 &        19195.549 &            0.003 \\[0.001mm]
   9 &        23036.530 &    8 &        21183.090 &            0.064 \\[0.001mm]
   9 &        23036.530 &    9 &        23055.314 &            1.828 \\[0.001mm]
   9 &        23036.530 &   10 &        24817.468 &            0.065 \\[0.001mm]
   9 &        23036.530 &   11 &        26474.473 &            0.005 \\[0.001mm]
  10 &        24799.750 &    8 &        21183.090 &            0.004 \\[0.001mm]
  10 &        24799.750 &    9 &        23055.314 &            0.072 \\[0.001mm]
  10 &        24799.750 &   10 &        24817.468 &            1.981 \\[0.001mm]
  10 &        24799.750 &   11 &        26474.473 &            0.072 \\[0.001mm]
  10 &        24799.750 &   12 &        28030.851 &            0.006 \\[0.001mm]
  11 &        26457.784 &    9 &        23055.314 &            0.005 \\[0.001mm]
  11 &        26457.784 &   10 &        24817.468 &            0.080 \\[0.001mm]
  11 &        26457.784 &   11 &        26474.473 &            2.143 \\[0.001mm]
  11 &        26457.784 &   12 &        28030.851 &            0.079 \\[0.001mm]
  11 &        26457.784 &   13 &        29490.540 &            0.008 \\[0.001mm]
  11 &        26457.784 &   14 &        30856.851 &            0.001 \\[0.001mm]
  12 &        28015.158 &   10 &        24817.468 &            0.007 \\[0.001mm]
  12 &        28015.158 &   11 &        26474.473 &            0.088 \\[0.001mm]
  12 &        28015.158 &   12 &        28030.851 &            2.315 \\[0.001mm]
  12 &        28015.158 &   13 &        29490.540 &            0.085 \\[0.001mm]
  12 &        28015.158 &   14 &        30856.851 &            0.009 \\[0.001mm]
  12 &        28015.158 &   15 &        32132.402 &            0.002 \\[0.001mm]
  13 &        29475.813 &   11 &        26474.473 &            0.008 \\[0.001mm]
  13 &        29475.813 &   12 &        28030.851 &            0.096 \\[0.001mm]
  13 &        29475.813 &   13 &        29490.540 &            2.498 \\[0.001mm]
  13 &        29475.813 &   14 &        30856.851 &            0.092 \\[0.001mm]
  13 &        29475.813 &   15 &        32132.402 &            0.011 \\[0.001mm]
  13 &        29475.813 &   16 &        33318.965 &            0.002 \\[0.001mm]
  14 &        30843.067 &   11 &        26474.473 &            0.001 \\[0.001mm]
  14 &        30843.067 &   12 &        28030.851 &            0.010 \\[0.001mm]
  14 &        30843.067 &   13 &        29490.540 &            0.104 \\[0.001mm]
  14 &        30843.067 &   14 &        30856.851 &            2.696 \\[0.001mm]
  14 &        30843.067 &   15 &        32132.402 &            0.098 \\[0.001mm]
  14 &        30843.067 &   16 &        33318.965 &            0.013 \\[0.001mm]
  14 &        30843.067 &   17 &        34416.761 &            0.003 \\[0.001mm]
  15 &        32119.545 &   12 &        28030.851 &            0.002 \\[0.001mm]
  15 &        32119.545 &   13 &        29490.540 &            0.012 \\[0.001mm]
  15 &        32119.545 &   14 &        30856.851 &            0.112 \\[0.001mm]
  15 &        32119.545 &   15 &        32132.402 &            2.912 \\[0.001mm]
  15 &        32119.545 &   16 &        33318.965 &            0.105 \\[0.001mm]
  15 &        32119.545 &   17 &        34416.761 &            0.014 \\[0.001mm]
  15 &        32119.545 &   18 &        35424.375 &            0.003 \\[0.001mm]
  16 &        33307.025 &   13 &        29490.540 &            0.002 \\[0.001mm]
  16 &        33307.025 &   14 &        30856.851 &            0.013 \\[0.001mm]
  16 &        33307.025 &   15 &        32132.402 &            0.120 \\[0.001mm]
  16 &        33307.025 &   16 &        33318.965 &            3.154 \\[0.001mm]
  16 &        33307.025 &   17 &        34416.761 &            0.113 \\[0.001mm]
  16 &        33307.025 &   18 &        35424.375 &            0.015 \\[0.001mm]
  16 &        33307.025 &   19 &        36341.801 &            0.004 \\[0.001mm]
  16 &        33307.025 &   20 &        37173.035 &            0.001 \\[0.001mm]
  17 &        34405.749 &   14 &        30856.851 &            0.003 \\[0.001mm]
  17 &        34405.749 &   15 &        32132.402 &            0.015 \\[0.001mm]
  17 &        34405.749 &   16 &        33318.965 &            0.130 \\[0.001mm]
  17 &        34405.749 &   17 &        34416.761 &            3.438 \\[0.001mm]
  17 &        34405.749 &   18 &        35424.375 &            0.124 \\[0.001mm]
  17 &        34405.749 &   19 &        36341.801 &            0.016 \\[0.001mm]
  17 &        34405.749 &   20 &        37173.035 &            0.006 \\[0.001mm]
  18 &        35414.305 &   15 &        32132.402 &            0.004 \\[0.001mm]
  18 &        35414.305 &   16 &        33318.965 &            0.016 \\[0.001mm]
  18 &        35414.305 &   17 &        34416.761 &            0.143 \\[0.001mm]
  18 &        35414.305 &   18 &        35424.375 &            3.791 \\[0.001mm]
  18 &        35414.305 &   19 &        36341.801 &            0.141 \\[0.001mm]
  18 &        35414.305 &   20 &        37173.035 &            0.016 \\[0.001mm]
  19 &        36332.666 &   16 &        33318.965 &            0.005 \\[0.001mm]
  19 &        36332.666 &   17 &        34416.761 &            0.017 \\[0.001mm]
  19 &        36332.666 &   18 &        35424.375 &            0.163 \\[0.001mm]
  19 &        36332.666 &   19 &        36341.801 &            4.234 \\[0.001mm]
  19 &        36332.666 &   20 &        37173.035 &            0.165 \\[0.001mm]
  20 &        37164.793 &   16 &        33318.965 &            0.001 \\[0.001mm]
  20 &        37164.793 &   17 &        34416.761 &            0.006 \\[0.001mm]
  20 &        37164.793 &   18 &        35424.375 &            0.017 \\[0.001mm]
  20 &        37164.793 &   19 &        36341.801 &            0.191 \\[0.001mm]
  20 &        37164.793 &   20 &        37173.035 &            4.775 \\[0.001mm]

\hline
 \end{longtable}
 \end{center}

%%%%%%%%%%%%%%%%%%%%%%%%%%%%%%%%%%%%%%%%%%%%%%%%%%%%%%%%%%%%%%%%%%%%%%%%%%%%%%%%%%%%%%%%%%%%%%%%%%%%%%%%%%%%%%%%%%%%%%%%

\begin{center}
\begin{longtable}{|cc|cc|c|}
\caption{\label{pi0-pi0-transitions} 
\rm{$^3\Pi_0 -^3\Pi_0$ transition
electric dipole matrix elements (in a.u.)
for  $J_A$=0,    $J'_A$=1.
The corresponding matrix elements for  $^3\Pi_1 -^3\Pi_1$ and $^3\Pi_2 -^3\Pi_2$ transitions
are nearly identical and are omitted for simplicity} }\\

\hline
\multicolumn{1}{|c}{} &\multicolumn{1}{c|}{} &\multicolumn{1}{|c}{} &\multicolumn{1}{c|}{} &\multicolumn{1}{c|}{} \\[2pt]
\multicolumn{1}{|c} { $v_A$}  & \multicolumn{1}{c|} {$E_{vJ}$(cm$^{-1}$)}      & 
 \multicolumn{1}{|c}{ $v'_A$} & \multicolumn{1}{c|} {$E_{v'J'}$(cm$^{-1}$)} &
 \multicolumn{1}{c|}{ $\vert M\vert^2$ (a.u.)} \\[2pt]
\multicolumn{1}{|c}{} &\multicolumn{1}{c|}{} &\multicolumn{1}{|c}{} &\multicolumn{1}{c|}{} &\multicolumn{1}{c|}{} \\[2pt]
\hline
\multicolumn{1}{|c}{} &\multicolumn{1}{c|}{} &\multicolumn{1}{|c}{} &\multicolumn{1}{c|}{} &\multicolumn{1}{c|}{} \\[2pt]
\endfirsthead

\hline
\multicolumn{5}{|c|} {  }\\[1pt]
 \multicolumn{5}{|c|} {\rm{ $^3\Pi_0 -^3\Pi_0$ transitions (continuation)} }\\[2pt]
\multicolumn{5}{|c|} {  }\\[1pt]
\hline
\multicolumn{1}{|c}{} &\multicolumn{1}{c|}{} &\multicolumn{1}{|c}{} &\multicolumn{1}{c|}{} &\multicolumn{1}{c|}{} \\[2pt]

\endhead

    0 &        29590.117 &    0 &        29616.792 &            1.309 \\[0.001mm]
   1 &        31568.468 &    1 &        31593.447 &            1.586 \\[0.001mm]
   2 &        33389.971 &    2 &        33413.313 &            1.917 \\[0.001mm]
   3 &        35060.018 &    3 &        35081.762 &            2.316 \\[0.001mm]
   3 &        35060.018 &    4 &        36601.163 &            0.131 \\[0.001mm]
   4 &        36581.006 &    3 &        35081.762 &            0.141 \\[0.001mm]
   4 &        36581.006 &    4 &        36601.163 &            2.812 \\[0.001mm]
   4 &        36581.006 &    5 &        37970.365 &            0.177 \\[0.001mm]
   5 &        37951.821 &    4 &        36601.163 &            0.191 \\[0.001mm]
   5 &        37951.821 &    5 &        37970.365 &            3.447 \\[0.001mm]
   5 &        37951.821 &    6 &        39183.750 &            0.230 \\[0.001mm]
   6 &        39166.901 &    5 &        37970.365 &            0.251 \\[0.001mm]
   6 &        39166.901 &    6 &        39183.750 &            4.302 \\[0.001mm]
   6 &        39166.901 &    7 &        40228.488 &            0.295 \\[0.001mm]
   7 &        40213.525 &    6 &        39183.750 &            0.324 \\[0.001mm]
   7 &        40213.525 &    7 &        40228.488 &            5.553 \\[0.001mm]
   7 &        40213.525 &    8 &        41076.108 &            0.368 \\[0.001mm]
   8 &        41063.494 &    7 &        40228.488 &            0.412 \\[0.001mm]
   8 &        41063.494 &    8 &        41076.108 &            7.717 \\[0.001mm]
   8 &        41063.494 &    9 &        41669.175 &            0.437 \\[0.001mm]
   9 &        41659.736 &    8 &        41076.108 &            0.505 \\[0.001mm]
   9 &        41659.736 &    9 &        41669.175 &           12.384 \\[0.001mm]
   9 &        41659.736 &   10 &        42018.670 &            0.531 \\[0.001mm]
  10 &        42011.891 &    9 &        41669.175 &            0.655 \\[0.001mm]
  10 &        42011.891 &   10 &        42018.670 &           20.115 \\[0.001mm]
  10 &        42011.891 &   12 &        42170.746 &            0.235 \\[0.001mm]
  11 &        42145.175 &   10 &        42018.670 &            0.105 \\[0.001mm]
  11 &        42145.175 &   11 &        42146.684 &           85.773 \\[0.001mm]
  11 &        42145.175 &   12 &        42170.746 &            1.858 \\[0.001mm]
  12 &        42168.395 &   10 &        42018.670 &            0.342 \\[0.001mm]
  12 &        42168.395 &   11 &        42146.684 &            3.633 \\[0.001mm]
  12 &        42168.395 &   12 &        42170.746 &           69.643 \\[0.001mm]

\hline
 \end{longtable}
 \end{center}
%%%%%%%%%%%%%%%%%%%%%%%%%%%%%%%%%%%%%%%%%%%%%%%%%%%%%%%%%%%%%%%%%%%%%%%%%%%%%%%%%%%%%%%%%%%%%%%%%%%%%%%%%%%%%%%%%%%%%%%%

%%%%%%%%%%%%%%%%%%%%%%%%%%%%%%%%%%%%%%%%%%%%%%%%%%%%%%%%%%%%%%%%%%%%%%%%%%%%%%%%%%%%%%%%%%%%%%%%%%%%%%%%%%%%%%%%%%%%%%%%

\begin{center}
\begin{longtable}{|cc|cc|c|}
\caption{\label{sigma-pi0-transitions}
\rm{ $^3\Sigma^-$ -  $^3\Pi_0$ transition
electric dipole matrix elements (in a.u.)
for  $N_x$=0, $J_X$=1 and  $J'_A$=0.
The corresponding matrix elements for  $^3\Sigma^-$ -$^3\Pi_1$ and $^3\Pi_2 -^3\Pi_2$ transitions
are nearly identical and are omitted for simplicity }}\\

\hline
\multicolumn{1}{|c}{} &\multicolumn{1}{c|}{} &\multicolumn{1}{|c}{} &\multicolumn{1}{c|}{} &\multicolumn{1}{c|}{} \\[2pt]
\multicolumn{1}{|c} { $v_X$}  & \multicolumn{1}{c|} {$E_{vNJ}$(cm$^{-1}$)}      & 
 \multicolumn{1}{|c}{ $v'_A$} & \multicolumn{1}{c|} {$E_{v'J'}$(cm$^{-1}$)} &
 \multicolumn{1}{c|}{ $\vert M\vert^2$ (a.u.)} \\[2pt]
\multicolumn{1}{|c}{} &\multicolumn{1}{c|}{} &\multicolumn{1}{|c}{} &\multicolumn{1}{c|}{} &\multicolumn{1}{c|}{} \\[2pt]
\hline
\multicolumn{1}{|c}{} &\multicolumn{1}{c|}{} &\multicolumn{1}{|c}{} &\multicolumn{1}{c|}{} &\multicolumn{1}{c|}{} \\[2pt]
\endfirsthead

\hline
\multicolumn{5}{|c|} {  }\\[1pt]
 \multicolumn{5}{|c|} {\rm{ $^3\Sigma^- -^3\Pi_0$ transitions (continuation)} }\\[2pt]
\multicolumn{5}{|c|} {  }\\[1pt]
\hline
\multicolumn{1}{|c}{} &\multicolumn{1}{c|}{} &\multicolumn{1}{|c}{} &\multicolumn{1}{c|}{} &\multicolumn{1}{c|}{} \\[2pt]

\endhead
   0 &         1540.906 &    0 &        29581.117 &            3.774 \\[0.001mm]
   0 &         1540.906 &    1 &        31559.468 &            2.364 \\[0.001mm]
   0 &         1540.906 &    2 &        33380.971 &            1.016 \\[0.001mm]
   0 &         1540.906 &    3 &        35051.018 &            0.389 \\[0.001mm]
   0 &         1540.906 &    4 &        36572.006 &            0.145 \\[0.001mm]
   0 &         1540.906 &    5 &        37942.821 &            0.055 \\[0.001mm]
   0 &         1540.906 &    6 &        39157.901 &            0.022 \\[0.001mm]
   0 &         1540.906 &    7 &        40204.525 &            0.009 \\[0.001mm]
   0 &         1540.906 &    8 &        41054.494 &            0.004 \\[0.001mm]
   0 &         1540.906 &    9 &        41650.736 &            0.002 \\[0.001mm]
   1 &         4499.658 &    0 &        29581.117 &            0.874 \\[0.001mm]
   1 &         4499.658 &    1 &        31559.468 &            0.736 \\[0.001mm]
   1 &         4499.658 &    2 &        33380.971 &            1.926 \\[0.001mm]
   1 &         4499.658 &    3 &        35051.018 &            1.599 \\[0.001mm]
   1 &         4499.658 &    4 &        36572.006 &            0.946 \\[0.001mm]
   1 &         4499.658 &    5 &        37942.821 &            0.491 \\[0.001mm]
   1 &         4499.658 &    6 &        39157.901 &            0.243 \\[0.001mm]
   1 &         4499.658 &    7 &        40204.525 &            0.119 \\[0.001mm]
   1 &         4499.658 &    8 &        41054.494 &            0.058 \\[0.001mm]
   1 &         4499.658 &    9 &        41650.736 &            0.025 \\[0.001mm]
   1 &         4499.658 &   10 &        42002.891 &            0.012 \\[0.001mm]
   1 &         4499.658 &   12 &        42159.395 &            0.002 \\[0.001mm]
   2 &         7301.650 &    0 &        29581.117 &            0.070 \\[0.001mm]
   2 &         7301.650 &    1 &        31559.468 &            0.883 \\[0.001mm]
   2 &         7301.650 &    2 &        33380.971 &            0.005 \\[0.001mm]
   2 &         7301.650 &    3 &        35051.018 &            0.766 \\[0.001mm]
   2 &         7301.650 &    4 &        36572.006 &            1.289 \\[0.001mm]
   2 &         7301.650 &    5 &        37942.821 &            1.158 \\[0.001mm]
   2 &         7301.650 &    6 &        39157.901 &            0.811 \\[0.001mm]
   2 &         7301.650 &    7 &        40204.525 &            0.504 \\[0.001mm]
   2 &         7301.650 &    8 &        41054.494 &            0.285 \\[0.001mm]
   2 &         7301.650 &    9 &        41650.736 &            0.139 \\[0.001mm]
   2 &         7301.650 &   10 &        42002.891 &            0.071 \\[0.001mm]
   2 &         7301.650 &   11 &        42136.175 &            0.002 \\[0.001mm]
   2 &         7301.650 &   12 &        42159.395 &            0.010 \\[0.001mm]
   3 &         9953.124 &    0 &        29581.117 &            0.002 \\[0.001mm]
   3 &         9953.124 &    1 &        31559.468 &            0.144 \\[0.001mm]
   3 &         9953.124 &    2 &        33380.971 &            0.514 \\[0.001mm]
   3 &         9953.124 &    3 &        35051.018 &            0.156 \\[0.001mm]
   3 &         9953.124 &    4 &        36572.006 &            0.087 \\[0.001mm]
   3 &         9953.124 &    5 &        37942.821 &            0.555 \\[0.001mm]
   3 &         9953.124 &    6 &        39157.901 &            0.819 \\[0.001mm]
   3 &         9953.124 &    7 &        40204.525 &            0.773 \\[0.001mm]
   3 &         9953.124 &    8 &        41054.494 &            0.568 \\[0.001mm]
   3 &         9953.124 &    9 &        41650.736 &            0.323 \\[0.001mm]
   3 &         9953.124 &   10 &        42002.891 &            0.178 \\[0.001mm]
   3 &         9953.124 &   11 &        42136.175 &            0.006 \\[0.001mm]
   3 &         9953.124 &   12 &        42159.395 &            0.026 \\[0.001mm]
   4 &        12460.153 &    1 &        31559.468 &            0.005 \\[0.001mm]
   4 &        12460.153 &    2 &        33380.971 &            0.171 \\[0.001mm]
   4 &        12460.153 &    3 &        35051.018 &            0.166 \\[0.001mm]
   4 &        12460.153 &    4 &        36572.006 &            0.300 \\[0.001mm]
   4 &        12460.153 &    5 &        37942.821 &            0.027 \\[0.001mm]
   4 &        12460.153 &    6 &        39157.901 &            0.067 \\[0.001mm]
   4 &        12460.153 &    7 &        40204.525 &            0.270 \\[0.001mm]
   4 &        12460.153 &    8 &        41054.494 &            0.361 \\[0.001mm]
   4 &        12460.153 &    9 &        41650.736 &            0.276 \\[0.001mm]
   4 &        12460.153 &   10 &        42002.891 &            0.177 \\[0.001mm]
   4 &        12460.153 &   11 &        42136.175 &            0.006 \\[0.001mm]
   4 &        12460.153 &   12 &        42159.395 &            0.028 \\[0.001mm]
   5 &        14828.776 &    2 &        33380.971 &            0.009 \\[0.001mm]
   5 &        14828.776 &    3 &        35051.018 &            0.138 \\[0.001mm]
   5 &        14828.776 &    4 &        36572.006 &            0.013 \\[0.001mm]
   5 &        14828.776 &    5 &        37942.821 &            0.211 \\[0.001mm]
   5 &        14828.776 &    6 &        39157.901 &            0.154 \\[0.001mm]
   5 &        14828.776 &    7 &        40204.525 &            0.021 \\[0.001mm]
   5 &        14828.776 &    8 &        41054.494 &            0.006 \\[0.001mm]
   5 &        14828.776 &    9 &        41650.736 &            0.030 \\[0.001mm]
   5 &        14828.776 &   10 &        42002.891 &            0.033 \\[0.001mm]
   5 &        14828.776 &   11 &        42136.175 &            0.001 \\[0.001mm]
   5 &        14828.776 &   12 &        42159.395 &            0.006 \\[0.001mm]
   6 &        17064.943 &    3 &        35051.018 &            0.009 \\[0.001mm]
   6 &        17064.943 &    4 &        36572.006 &            0.076 \\[0.001mm]
   6 &        17064.943 &    5 &        37942.821 &            0.010 \\[0.001mm]
   6 &        17064.943 &    6 &        39157.901 &            0.055 \\[0.001mm]
   6 &        17064.943 &    7 &        40204.525 &            0.134 \\[0.001mm]
   6 &        17064.943 &    8 &        41054.494 &            0.096 \\[0.001mm]
   6 &        17064.943 &    9 &        41650.736 &            0.037 \\[0.001mm]
   6 &        17064.943 &   10 &        42002.891 &            0.013 \\[0.001mm]
   6 &        17064.943 &   12 &        42159.395 &            0.001 \\[0.001mm]
   7 &        19174.511 &    4 &        36572.006 &            0.006 \\[0.001mm]
   7 &        19174.511 &    5 &        37942.821 &            0.024 \\[0.001mm]
   7 &        19174.511 &    6 &        39157.901 &            0.035 \\[0.001mm]
   7 &        19174.511 &    8 &        41054.494 &            0.026 \\[0.001mm]
   7 &        19174.511 &    9 &        41650.736 &            0.038 \\[0.001mm]
   7 &        19174.511 &   10 &        42002.891 &            0.029 \\[0.001mm]
   7 &        19174.511 &   11 &        42136.175 &            0.001 \\[0.001mm]
   7 &        19174.511 &   12 &        42159.395 &            0.005 \\[0.001mm]
   8 &        21163.200 &    4 &        36572.006 &            0.002 \\[0.001mm]
   8 &        21163.200 &    5 &        37942.821 &            0.001 \\[0.001mm]
   8 &        21163.200 &    6 &        39157.901 &            0.003 \\[0.001mm]
   8 &        21163.200 &    7 &        40204.525 &            0.024 \\[0.001mm]
   8 &        21163.200 &    8 &        41054.494 &            0.015 \\[0.001mm]
   8 &        21163.200 &    9 &        41650.736 &            0.002 \\[0.001mm]
   9 &        23036.530 &    4 &        36572.006 &            0.002 \\[0.001mm]
   9 &        23036.530 &    5 &        37942.821 &            0.007 \\[0.001mm]
   9 &        23036.530 &    8 &        41054.494 &            0.003 \\[0.001mm]
   9 &        23036.530 &    9 &        41650.736 &            0.007 \\[0.001mm]
   9 &        23036.530 &   10 &        42002.891 &            0.005 \\[0.001mm]
  10 &        24799.750 &    5 &        37942.821 &            0.006 \\[0.001mm]
  10 &        24799.750 &    6 &        39157.901 &            0.016 \\[0.001mm]
  11 &        26457.784 &    5 &        37942.821 &            0.002 \\[0.001mm]
  11 &        26457.784 &    6 &        39157.901 &            0.018 \\[0.001mm]
  11 &        26457.784 &    7 &        40204.525 &            0.024 \\[0.001mm]
  12 &        28015.158 &    6 &        39157.901 &            0.007 \\[0.001mm]
  12 &        28015.158 &    7 &        40204.525 &            0.041 \\[0.001mm]
  12 &        28015.158 &    8 &        41054.494 &            0.013 \\[0.001mm]
  12 &        28015.158 &    9 &        41650.736 &            0.009 \\[0.001mm]
  12 &        28015.158 &   10 &        42002.891 &            0.002 \\[0.001mm]
  13 &        29475.813 &    6 &        39157.901 &            0.001 \\[0.001mm]
  13 &        29475.813 &    7 &        40204.525 &            0.023 \\[0.001mm]
  13 &        29475.813 &    8 &        41054.494 &            0.054 \\[0.001mm]
  13 &        29475.813 &    9 &        41650.736 &            0.002 \\[0.001mm]
  13 &        29475.813 &   10 &        42002.891 &            0.001 \\[0.001mm]
  14 &        30843.067 &    7 &        40204.525 &            0.007 \\[0.001mm]
  14 &        30843.067 &    8 &        41054.494 &            0.058 \\[0.001mm]
  14 &        30843.067 &    9 &        41650.736 &            0.012 \\[0.001mm]
  14 &        30843.067 &   10 &        42002.891 &            0.016 \\[0.001mm]
  14 &        30843.067 &   12 &        42159.395 &            0.002 \\[0.001mm]
  15 &        32119.545 &    7 &        40204.525 &            0.001 \\[0.001mm]
  15 &        32119.545 &    8 &        41054.494 &            0.033 \\[0.001mm]
  15 &        32119.545 &    9 &        41650.736 &            0.052 \\[0.001mm]
  15 &        32119.545 &   10 &        42002.891 &            0.015 \\[0.001mm]

\hline
 \end{longtable}
 \end{center}

\section{APPENDIX C: He-OH$^+(X)$ potential energy surface}

The He-OH$^+$($^3\Sigma^-$)  interaction potential has being calculated
on a grid of points in Jacobi coordinates, where
$r$ is the intenuclear OH$^+$ distance, $R$ is the
distance between He and the 
center of mass of the OH$^+$, and  $\theta$ is the angle 
 between the vectors of lengths $R$ and $r$ (with
 $\theta=0^{\circ}$ corresponding to the He-H-O collinear 
arrangement).  
{\em Ab initio} calculations are performed
for 57 intermolecular
 $R$ distances in the range between 2.75 a$_0$ and 32 $a_0$
 distributed in the vicinity of the minimum
with a step of 0.1 a$_0$ and 0.25 or 0.5 a$_0$ outside. The angular 
$\theta$ coordinate is represented by a grid of 15 Gauss-Legendre nodes.
 The intramolecular distance $r$ is varied between 1.7 a$_0$ and 2.6 a$_0$
on a grid of 5 radial points, which is enough to describe the first 
vibrational levels of OH$^+$($^3\Sigma^-$).

We use the spin unrestricted coupled cluster method with single, 
double and non-iterative triple excitations 
(UCCSD(T))~\citep{knowles:93,knowles:00} for the calculations 
of total energies of the ground electronic state of He-OH$^+$, using 
  MOLPRO  program~\citep{MOLPRO}. 
 The system is well described
 by a single-determinant wave function, therefore we use 
restricted Hartree-Fock (RHF) calculations
as a reference for subsequent UCCSD(T) calculations.
 We used the augmented correlation-consistent quadruple-zeta (AVQZ) basis 
set~\citep{Dunning:89} of Dunning {\em et al.} augmented with 
 $3s2p1d$ mid-bond functions ($s: 0.553063,0.250866,0.117111$,
$p:0.392,0.142$, $d:0.328$).  
The standard Boys and Bernardi \citep{boys:70} counterpoise procedure
 is used to correct the interaction energy for the basis 
set superposition error
(BSSE).

We have used the UCCSD(T) method for the He-OH+ PES instead of MRCI as 
before for the sole diatom as it avoids the size-consistency problems and
 recovers larger portion of correlation energy due to the perturbative
inclusion of triple excitations, which is important for van der Waals complexes
 containing helium. The choice of AVQZ+bond functions basis offers good ratio of 
accuracy (energies can be close to the complete basis set limit) to computational 
time, as we have more degrees of freedom in comparison to sole OH cation.

The potential is expanded  in a series of Legendre polynomials, 
$V(R,r,\theta)=\sum_{l=0}^{15}v_l(R,r)P_l(\cos \theta)$ 
in order to represent the potential in the analytical form. The
radial expansion coefficients for each $r$ are interpolated 
along intermolecular distance $R$ using Reproducing 
Kernel Hilbert space method. The dependence on the $r$ intramolecular
distance is obtained by polynomial expansion 
in a dimensionally reduced $z=(r-r_e)/r_e$ coordinate.

We averaged the three-dimensional (3-D) $V(R, r, \theta)$ PES over the ground
 vibrational state of the OH$^+$ cation to obtain the $V_0(R,\theta)$ He-OH$^+$ potential that 
we can use in subsequent scattering calculations.  
The potential exhibits global minimum at collinear He-H-O geometry
for $\theta=0^{\circ}$. The well depth of the $V_0(R,\theta)$  potential
 is $D_e=729.6$ cm$^{-1}$ located at $R_e=4.79$ a$_0$. These values can be
 compared to potential published almost two decades
before by Meuwly and coworkers~\cite{meuwly:98}. Meuwly {\em et al.} 
reports $D_e$ of 701 cm$^{-1}$ and $R_e$ of 4.83 a$_0$. These are quite similar
 values in comparison to result in this work
 with our potentials being slightly more attractive. 
The origin of the deep well for this helium complex lies in the fact 
that OH$^+$($^3\Sigma^-$)'s positive charge acts as an acceptor of the
 helium electron cloud acting as a donor. This simple 
 model is discussed by Hughes {\em et al.} in their studies 
of HeOH$^+$ molecule (~\cite{hughes:97}). The He atom binds strongly
 to the protonated side of the OH$^+$ in the entrance channel of reaction to 
form the strongly bound molecular HeH$^+$ ion. 
 The zero-point dissociation energy of the He-OH$^+$ complex is 
reported(~\cite{meuwly:98}) to be around 360 cm$^{-1}$. The $D_0$ value 
calculated with our new UCCSD(T)  potential is 391 cm$^{-1}$. 
This indicates a fair stability of this helium complex.     

\section{APPENDIX D: He+OH$^+$ inelastic collisions}

As described above, in the OH$^+$($X ^{3}\Sigma^{-}$) electronic ground state,
 the rotational levels are split by spin-rotation coupling as previously mentioned in section 2, and
 the rotational wave functions 
 written for $J\geq 1$ as \citep{Gordy:84,lique:05}:
\begin{eqnarray} \label{F}
|F_{1}JM\rangle & = & 
\cos\alpha|N=J-1,SJM\rangle \nonumber \\
 & & +\sin\alpha|N=J+1,SJM\rangle \nonumber \\
|F_{2}JM\rangle & = & |N=J,SJM\rangle\\
|F_{3}JM\rangle & = & 
-\sin\alpha|N=J-1,SJM\rangle \nonumber \\
 & & +\cos\alpha|N=J+1,SJM\rangle \nonumber
\end{eqnarray}
where $|N,SJM\rangle$ denotes pure Hund's case (b) basis functions (see Appendix B) and
 the mixing
 angle $\alpha$ is obtained by diagonalization of the molecular Hamiltonian. 
 In the pure case (b) limit, $\alpha 
\rightarrow 0$, the $F_{1}$ level corresponds to $N=J-1$ and the $F_{3}$ level to $N=J+1$. 

The rotational energy
levels of the OH$^+$ molecule were computed with the use of experimental
spectroscopic constants of \citep{Merer-etal:75}.

The quantal coupled equations were solved in the intermediate coupling scheme 
using the MOLSCAT code \citep{molscat:94} modified to take 
into account the fine structure of the energy levels. 

For the OH$^+$ molecule, an additional splitting of rotational levels exists. The Hydrogen atom
possesses a non-zero nuclear spin ($I=0.5$) so that the energy levels of OH$^+$ are
characterized by the quantum numbers $N$, $J$ and $F$, where $F$
results from the coupling of $\vec{J}$ with $\vec{I}$ ($\vec{F} =
\vec{J} + \vec{I}$).
The hyperfine splitting of the OH$^+$ levels being very small, 
the hyperfine levels can be safely assumed to be degenerate as 
was considered in the transitions treated in the previous section. Then, 
it is possible to simplify considerably the hyperfine scattering problem.
The integral cross sections corresponding to transitions between hyperfine 
levels of the OH$^+$ molecules can be obtained from scattering S-matrix between
 fine structure levels using a recoupling method \citep{Alexander:85hyp}. 

Inelastic cross sections associated with a transition from an initial hyperfine
 level $N,J,F$ to a hyperfine level $N',J',F'$ were thus obtained as follow :

\begin{eqnarray}
\sigma_{NJF \to N'J'F'}  & = & \frac{\pi}{k^{2}_{NJF}} (2F'+1) \sum_{K} \nonumber \\
& & \times \left\{ 
\begin{array}{ccc}
J & J' & K \\
F' & F & I 
\end{array}
\right\}^2
P^K(J \to J')
\end{eqnarray}
The $P^{K}(J \to J')$ are the tensor opacities defined by  :
\begin{equation}
P^{K}(J \to J')=\frac{1}{2K+1}\sum_{ll'}|T^{K}(Jl;J'l')|^{2}
\end{equation}
The reduced T-matrix elements (where $T = 1 - S$) are defined by \citep{alexander:83}: 
\begin{eqnarray}
T^{K}(Jl;J'l') & = & (-1)^{-J-l'}(2K+1)\sum_{J_t}(-1)^{J_t}(2J_t+1) \nonumber\\
& & \times  \left\{\begin{array}{ccc}
l' & J' & J_t \\ 
J & l & K 
\end{array}\right\}
T^{J_t}(Jl;J'l')
\end{eqnarray}
 where $J_t=J+l$ is the total triatomic angular momentum, and $l$
is the orbital angular momentum quantum number.

The
scattering calculations were carried out on  total energy, $E_{tot}$, grid with a variable
steps. For the energies below 1000~cm$^{-1}$ the step was equal to
1~cm$^{-1}$, then, between 1000 and 1500~cm$^{-1}$ it was increased to
10~cm$^{-1}$, and to 100~cm$^{-1}$ for energy interval from 1500 to
2200 cm$^{-1}$.
In order to ensure convergence of the inelastic cross
sections, it is necessary to include in the calculations several
energetically inaccessible (closed) levels.  At the largest energies
considered in this work, the OH$^+$ rotational basis were extended
to $N=10$ to ensure convergence of the cross sections of
OH$^+$.

\newpage
%\bibliography{joint-biblio}

\begin{thebibliography}{96}
\expandafter\ifx\csname natexlab\endcsname\relax\def\natexlab#1{#1}\fi

\bibitem[{Ag\'undez \& Cernicharo(2006)}]{Agundez-Cernicharo:06}
Ag\'undez, M., \& Cernicharo, J. 2006, ApJ, 650, 374.

\bibitem[{{Ag{\'u}ndez} {et~al.}(2008){Ag{\'u}ndez}, {Fonfr{\'{\i}}a},
  {Cernicharo}, {Pardo}, \& {Gu{\'e}lin}}]{Agundez2008}
{Ag{\'u}ndez}, M., {Fonfr{\'{\i}}a}, J.~P., {Cernicharo}, J., {Pardo}, J.~R.,
  \& {Gu{\'e}lin}, M. 2008, \aap, 479, 493.

\bibitem[{{Aleman} {et~al.}(2014){Aleman}, {Ueta}, {Ladjal}, {Exter},
  {Kastner}, {Montez}, {Tielens}, {Chu}, {Izumiura}, {McDonald}, {Sahai},
  {Siodmiak}, {Szczerba}, {van Hoof}, {Villaver}, {Vlemmings}, {Wittkowski}, \&
  {Zijlstra}}]{Aleman-etal:14}
{Aleman}, I., {Ueta}, T., {Ladjal}, D., {et~al.} 2014, ArXiv e-prints,
  arXiv:1404.2431.

\bibitem[{Alexander(1985)}]{alexander:85}
Alexander, M.~H. 1985, Chem. Phys., 92, 337.

\bibitem[{{Alexander} \& {Dagdigian}(1983)}]{alexander:83}
{Alexander}, M.~H., \& {Dagdigian}, P.~J. 1983, \jcp, 79, 302.

\bibitem[{{Alexander} \& {Dagdigian}(1985)}]{Alexander:85hyp}
---. 1985, \jcp, 83, 2191.

\bibitem[{Aslan {et~al.}(2012)Aslan, Bulut, Castillo, Ba{\~ n}ares, Aoiz, \&
  Roncero}]{Aslan-etal:12}
Aslan, E., Bulut, N., Castillo, J.~F., {et~al.} 2012, ApJ, 739, 31.

\bibitem[{{Benz} {et~al.}(2010){Benz}, {Bruderer}, {van Dishoeck},
  {St{\"a}uber}, {Wampfler}, {Melchior}, {Dedes}, {Wyrowski}, {Doty}, {van der
  Tak}, {B{\"a}chtold}, {Csillaghy}, {Megej}, {Monstein}, {Soldati},
  {Bachiller}, {Baudry}, {Benedettini}, {Bergin}, {Bjerkeli}, {Blake},
  {Bontemps}, {Braine}, {Caselli}, {Cernicharo}, {Codella}, {Daniel}, {di
  Giorgio}, {Dieleman}, {Dominik}, {Encrenaz}, {Fich}, {Fuente}, {Giannini},
  {Goicoechea}, {de Graauw}, {Helmich}, {Herczeg}, {Herpin}, {Hogerheijde},
  {Jacq}, {Jellema}, {Johnstone}, {J{\o}rgensen}, {Kristensen}, {Larsson},
  {Lis}, {Liseau}, {Marseille}, {McCoey}, {Melnick}, {Neufeld}, {Nisini},
  {Olberg}, {Ossenkopf}, {Parise}, {Pearson}, {Plume}, {Risacher},
  {Santiago-Garc{\'{\i}}a}, {Saraceno}, {Schieder}, {Shipman}, {Stutzki},
  {Tafalla}, {Tielens}, {van Kempen}, {Visser}, \& {Y{\i}ld{\i}z}}]{Benz2010}
{Benz}, A.~O., {Bruderer}, S., {van Dishoeck}, E.~F., {et~al.} 2010, \aap, 521,
  L35.

\bibitem[{Boys \& Bernardi(1970)}]{boys:70}
Boys, S.~F., \& Bernardi, F. 1970, Mol. Phys., 19, 553.

\bibitem[{Brzozowski {et~al.}(1974)Brzozowski, Elander, Erman, \&
  Lyyra}]{Brzozowski-etal:74}
Brzozowski, J., Elander, N., Erman, P., \& Lyyra, M. 1974, Phys. Scripta, 10,
  241.

\bibitem[{Burley {et~al.}(1987)Burley, Ervin, \& Armentrout}]{Burley-etal:87}
Burley, J.~D., Ervin, K.~M., \& Armentrout, P. 1987, Int. J. Mass Spect. and
  Ion Processes, 80, 153.

\bibitem[{{Cernicharo}(2012)}]{Cernicharo2012}
{Cernicharo}, J. 2012, in Proceedings of the European Conference on Laboratory
  Astrophysics, European Astronomical Society Publications Series, ed.
  C.~{Stehl{\'e}}, C.~{Joblin}, \& L.~{d'Hendecourt}

\bibitem[{{Cernicharo} {et~al.}(2010){Cernicharo}, {Decin}, {Barlow},
  {Ag{\'u}ndez}, {Royer}, {Vandenbussche}, {Wesson}, {Polehampton}, {De Beck},
  {Blommaert}, {Daniel}, {De Meester}, {Exter}, {Feuchtgruber}, {Gear},
  {Goicoechea}, {Gomez}, {Groenewegen}, {Hargrave}, {Huygen}, {Imhof},
  {Ivison}, {Jean}, {Kerschbaum}, {Leeks}, {Lim}, {Matsuura}, {Olofsson},
  {Posch}, {Regibo}, {Savini}, {Sibthorpe}, {Swinyard}, {Vandenbussche}, \&
  {Waelkens}}]{Cernicharo2010}
{Cernicharo}, J., {Decin}, L., {Barlow}, M.~J., {et~al.} 2010, \aap, 518, L136.

\bibitem[{Chen \& Guo(1996)}]{Chen-Guo:96}
Chen, R., \& Guo, H. 1996, \jcp, 105, 3569.

\bibitem[{Christoffel \& Bowman(1983)}]{christoffel:90}
Christoffel, K.~M., \& Bowman, J.~M. 1983, \jcp, 78, 3952.

\bibitem[{{D. E. Woon and Jr. T. H. Dunning}(1994)}]{Woon-Duninng:94}
{D. E. Woon and Jr. T. H. Dunning}. 1994, \jcp, 101, 8877.

\bibitem[{Davidson(1975)}]{Davidson:75}
Davidson, E.~R. 1975, J. Comp. Phys., 17, 87.

\bibitem[{de~Almeida \& {Singh}(1981)}]{Almeida-Singh:81}
de~Almeida, A.~A., \& {Singh}, P.~D. 1981, ApJL, 95,
  383, 383

\bibitem[{{Dubernet, M.-L.} {et~al.}(2013){Dubernet, M.-L.}, {Alexander, M.
  H.}, {Ba, Y. A.}, {Balakrishnan, N.}, {Balan\c{c}a, C.}, {Ceccarelli, C.},
  {Cernicharo, J.}, {Daniel, F.}, {Dayou, F.}, {Doronin, M.}, {Dumouchel, F.},
  {Faure, A.}, {Feautrier, N.}, {Flower, D. R.}, {Grosjean, A.}, {Halvick, P.},
  {Klos, J.}, {Lique, F.}, {McBane, G. C.}, {Marinakis, S.}, {Moreau, N.},
  {Moszynski, R.}, {Neufeld, D. A.}, {Roueff, E.}, {Schilke, P.}, {Spielfiedel,
  A.}, {Stancil, P. C.}, {Stoecklin, T.}, {Tennyson, J.}, {Yang, B.},
  {Vasserot, A.-M.}, \& {Wiesenfeld, L.}}]{Dubernet:13}
{Dubernet, M.-L.}, {Alexander, M. H.}, {Ba, Y. A.}, {et~al.} 2013, \aap, 553,
  A50.

\bibitem[{{Dumouchel} {et~al.}(2012){Dumouchel}, {Spielfiedel}, {Senent}, \&
  {Feautrier}}]{Dumouchel:12}
{Dumouchel}, F., {Spielfiedel}, A., {Senent}, M.~L., \& {Feautrier}, N. 2012,
  Chem. Phys. Lett., 533, 6.

\bibitem[{Dunning \& Jr.(1989)}]{Dunning:89}
Dunning, T.~H., \& Jr. 1989, \jcp, 90, 1007.

\bibitem[{{Etxaluze} {et~al.}(2013){Etxaluze}, {Goicoechea}, {Cernicharo},
  {Polehampton}, {Noriega-Crespo}, {Molinari}, {Swinyard}, {Wu}, \&
  {Bally}}]{Etxaluze-etal:13}
{Etxaluze}, M., {Goicoechea}, J.~R., {Cernicharo}, J., {et~al.} 2013, \aap,
  556, A137.

\bibitem[{{Etxaluze} {et~al.}(2014){Etxaluze}, {Cernicharo}, {Goicoechea}, {van
  Hoof}, {Swinyard}, {Barlow}, {van de Steene}, {Groenewegen}, {Kerschbaum},
  {Lim}, {Lique}, {Matsuura}, {Pearson}, {Polehampton}, {Royer}, \&
  {Ueta}}]{Etxaluze-etal:14}
{Etxaluze}, M., {Cernicharo}, J., {Goicoechea}, J.~R., {et~al.} 2014, ArXiv
  e-prints, arXiv:1404.2177

\bibitem[{{Faure} \& {Tennyson}(2001)}]{Faure2001}
{Faure}, A., \& {Tennyson}, J. 2001, \mnras, 325, 443.

\bibitem[{Gerin {et~al.}(2010)Gerin, {De Luca, M.}, {Black, J.}, {Goicoechea,
  J. R.}, {Herbst, E.}, {Neufeld, D. A.}, {Falgarone, E.}, {Godard, B.},
  {Pearson, J. C.}, {Lis, D. C.}, {Phillips, T. G.}, {Bell, T. A.},
  {Sonnentrucker, P.}, {Boulanger, F.}, {Cernicharo, J.}, {Coutens, A.},
  {Dartois, E.}, {Encrenaz, P.}, {Giesen, T.}, {Goldsmith, P. F.}, {Gupta, H.},
  {Gry, C.}, {Hennebelle, P.}, {Hily-Blant, P.}, {Joblin, C.}, {Kazmierczak,
  M.}, {Kolos, R.}, {Krelowski, J.}, {Martin-Pintado, J.}, {Monje, R.},
  {Mookerjea, B.}, {Perault, M.}, {Persson, C.}, {Plume, R.}, {Rimmer, P. B.},
  {Salez, M.}, {Schmidt, M.}, {Stutzki, J.}, {Teyssier, D.}, {Vastel, C.}, {Yu,
  S.}, {Contursi, A.}, {Menten, K.}, {Geballe, T.}, {Schlemmer, S.}, {Shipman,
  R.}, {Tielens, A. G. G. M.}, {Philipp-May, S.}, {Cros, A.}, {Zmuidzinas, J.},
  {Samoska, L. A.}, {Klein, K.}, \& {Lorenzani, A.}}]{Gerin-etal:10}
Gerin, M., {De Luca, M.}, {Black, J.}, {et~al.} 2010, \aap, 518,
  L110.

\bibitem[{Gioumousis \& Stevenson(1958)}]{Gioumousis-Stevenson:58}
Gioumousis, G., \& Stevenson, D.~P. 1958, \jcp, 29, 294.

\bibitem[{{Godard} \& {Cernicharo}(2013)}]{godard:13}
{Godard}, B., \& {Cernicharo}, J. 2013, \aap, 550, A8.

\bibitem[{{Godard} {et~al.}(2012){Godard}, {Falgarone}, {Gerin}, {Lis}, {De
  Luca}, {Black}, {Goicoechea}, {Cernicharo}, {Neufeld}, {Menten}, \&
  {Emprechtinger}}]{Godard2012}
{Godard}, B., {Falgarone}, E., {Gerin}, M., {et~al.} 2012, \aap, 540, A87.

\bibitem[{{Goicoechea} {et~al.}(2013){Goicoechea}, {Etxaluze}, {Cernicharo},
  {Gerin}, {Neufeld}, {Contursi}, {Bell}, {De Luca}, {Encrenaz}, {Indriolo},
  {Lis}, {Polehampton}, \& {Sonnentrucker}}]{Goicoechea-etal:13}
{Goicoechea}, J.~R., {Etxaluze}, M., {Cernicharo}, J., {et~al.} 2013, \apjl,
  769, L13.

\bibitem[{Goldfield {et~al.}(1995)Goldfield, Gray, \&
  Schatz}]{Goldfield-etal:95}
Goldfield, E.~M., Gray, S.~K., \& Schatz, G.~C. 1995, \jcp, 102,
  8807.

\bibitem[{G{\'o}mez-Carrasco \& Roncero(2006)}]{Gomez-Carrasco-Roncero:06}
G{\'o}mez-Carrasco, S., \& Roncero, O. 2006, \jcp, 125, 054102.

\bibitem[{{Gonz{\'a}lez-Alfonso} \& {Cernicharo}(1999)}]{Gonzalez-Alfonso1999}
{Gonz{\'a}lez-Alfonso}, E., \& {Cernicharo}, J. 1999, in ESA Special
  Publication, Vol. 427, The Universe as Seen by ISO, ed. P.~{Cox} \&
  M.~{Kessler}, 325.

\bibitem[{Gonz{\'a}lez-Alfonso {et~al.}(2013)Gonz{\'a}lez-Alfonso, {Fischer,
  J.}, {Bruderer, S.}, {Müller, H. S. P.}, {Graci{\'a}-Carpio, J.}, {Sturm,
  E.}, {Lutz, D.}, {Poglitsch, A.}, {Feuchtgruber, H.}, {Veilleux, S.},
  {Contursi, A.}, {Sternberg, A.}, {Hailey-Dunsheath, S.}, {Verma, A.},
  {Christopher, N.}, {Davies, R.}, {Genzel, R.}, \& {Tacconi,
  L.}}]{Gonzalez-Alfonso-etal:13}
Gonz{\'a}lez-Alfonso, E., {Fischer, J.}, {Bruderer, S.}, {et~al.} 2013, \aap, 550, A25

\bibitem[{Gonz\'alez-Lezana {et~al.}(2005)Gonz\'alez-Lezana, Aguado, Paniagua,
  \& Roncero}]{Gonzalez-Lezana-etal:05}
Gonz\'alez-Lezana, T., Aguado, A., Paniagua, M., \& Roncero, O. 2005, \jcp, 123, 194309.

\bibitem[{Gordy \& Cook(1984)}]{Gordy:84}
Gordy, W., \& Cook, R.~L. 1984, Microwave molecular spectra, Wileys and sons.

\bibitem[{Gray \& Balint-Kurti(1998)}]{Gray-Balint-Kurti:98}
Gray, S.~K., \& Balint-Kurti, G.~G. 1998, \jcp, 108, 950.

\bibitem[{Gruebele {et~al.}(1986)Gruebele, M\"uler, \&
  Saykally}]{Gruebele-etal:86}
Gruebele, M. H.~W., M\"uler, R.~P., \& Saykally, R.~J. 1986, \jcp,  84, 2489.

\bibitem[{Herzberg(1950)}]{Herzberg-diatomics}
Herzberg, G. 1950, Molecular spectra and molecular structure. I. Spectra of
  diatomic molecules, van Nostrand Reinhold Co. (New York).

\bibitem[{{Hily-Blant} {et~al.}(2010){Hily-Blant}, {Maret}, {Bacmann},
  {Bottinelli}, {Parise}, {Caux}, {Faure}, {Bergin}, {Blake}, {Castets},
  {Ceccarelli}, {Cernicharo}, {Coutens}, {Crimier}, {Demyk}, {Dominik},
  {Gerin}, {Hennebelle}, {Henning}, {Kahane}, {Klotz}, {Melnick}, {Pagani},
  {Schilke}, {Vastel}, {Wakelam}, {Walters}, {Baudry}, {Bell}, {Benedettini},
  {Boogert}, {Cabrit}, {Caselli}, {Codella}, {Comito}, {Encrenaz}, {Falgarone},
  {Fuente}, {Goldsmith}, {Helmich}, {Herbst}, {Jacq}, {Kama}, {Langer},
  {Lefloch}, {Lis}, {Lord}, {Lorenzani}, {Neufeld}, {Nisini}, {Pacheco},
  {Phillips}, {Salez}, {Saraceno}, {Schuster}, {Tielens}, {van der Tak}, {van
  der Wiel}, {Viti}, {Wyrowski}, \& {Yorke}}]{Hily-Blant2010}
{Hily-Blant}, P., {Maret}, S., {Bacmann}, A., {et~al.} 2010, \aap, 521, L52.

\bibitem[{Hollenbach {et~al.}(2012)Hollenbach, Kaufman, Neufeld, Wolfire, \&
  Goicoechea}]{Hollenbach-etal:12}
Hollenbach, D., Kaufman, M.~J., Neufeld, D., Wolfire, M., \& Goicoechea, J.~R.
  2012,\apj, 105, 754.

\bibitem[{Huang {et~al.}(1996)Huang, Iyengar, Kouri, \&
  Hoffman}]{Huang-etal:96}
Huang, Y., Iyengar, S.~S., Kouri, D.~J., \& Hoffman, D.~K. 1996, \jcp, 105, 927.

\bibitem[{Huang {et~al.}(1994)Huang, Kouri, \& Hoffman}]{Huang-etal:94}
Huang, Y., Kouri, D.~J., \& Hoffman, D.~K. 1994, \jcp, 101, 10493.

\bibitem[{Hughes \& von Nagy-Felsobuki(1997)}]{hughes:97}
Hughes, J.~M., \& von Nagy-Felsobuki, E.~I. 1997, J. Phys. Chem. A, 101, 3995.

\bibitem[{Hutson \& Green(1994)}]{molscat:94}
Hutson, J.~M., \& Green, S. 1994, {\sc molscat} computer code, version 14
  (1994), distributed by Collaborative Computational Project No. 6 of the
  Engineering and Physical Sciences Research Council (UK)

\bibitem[{{Indriolo} {et~al.}(2007){Indriolo}, {Geballe}, {Oka}, \&
  {McCall}}]{Indriolo2007}
{Indriolo}, N., {Geballe}, T.~R., {Oka}, T., \& {McCall}, B.~J. 2007, \apj,
  671, 1736.

\bibitem[{{Indriolo} \& {McCall}(2012)}]{Indriolo2012}
{Indriolo}, N., \& {McCall}, B.~J. 2012, \apj, 745, 91.

\bibitem[{Knowles {et~al.}(1993)Knowles, Hampel, \& Werner}]{knowles:93}
Knowles, P.~J., Hampel, C., \& Werner, H.-J. 1993, \jcp, 99, 5219.

\bibitem[{Knowles {et~al.}(2000)Knowles, Hampel, \& Werner}]{knowles:00}
---. 2000, \jcp, 112, 3106.

\bibitem[{Krelowski {et~al.}(2010)Krelowski, Beletsky, \&
  Galazutdinov}]{Krelowski-etal:10}
Krelowski, J., Beletsky, Y., \& Galazutdinov, G.~A. 2010, \apjl, 719, L20.

\bibitem[{Kroes \& Neuhauser(1996)}]{Kroes-Neuhauser:96}
Kroes, G.~J., \& Neuhauser, D. 1996, \jcp, 105, 8690.

\bibitem[{Langevin(1905)}]{Langevin:1905}
Langevin, P. 1905, Ann. Chim. Phys., 5, 245.

\bibitem[{{Larsson}(1983)}]{Larsson1983}
{Larsson}, M. 1983, \aap, 128, 291.

\bibitem[{{Le Bourlot} {et~al.}(2012){Le Bourlot}, {Le Petit}, {Pinto},
  {Roueff}, \& {Roy}}]{Le-Bourlot2012}
{Le Bourlot}, J., {Le Petit}, F., {Pinto}, C., {Roueff}, E., \& {Roy}, F. 2012,
  \aap, 541, A76.

\bibitem[{{Le Petit} {et~al.}(2006){Le Petit}, {Nehm{\'e}}, {Le Bourlot}, \&
  {Roueff}}]{Le-Petit2006}
{Le Petit}, F., {Nehm{\'e}}, C., {Le Bourlot}, J., \& {Roueff}, E. 2006, \apjs,
  164, 506, 506

\bibitem[{Lefebvre-Brion \& Field(1986)}]{Lefebvre-Brion-Field:86}
Lefebvre-Brion, H., \& Field, R.~W. 1986, Perturbations in the Spectra of
  Diatomic Molecules, Academic Press, London.

\bibitem[{{Lique}(2010)}]{lique:10}
{Lique}, F. 2010, \jcp, 132, 044311.

\bibitem[{{Lique} \& {K{\l}os}(2011)}]{lique:11CN}
{Lique}, F., \& {K{\l}os}, J. 2011, \mnras, 413, L20.

\bibitem[{Lique {et~al.}(2005)Lique, Spielfiedel, Dubernet, \&
  Feautrier}]{lique:05}
Lique, F., Spielfiedel, A., Dubernet, M.~L., \& Feautrier, N. 2005, \jcp, 123,
  134316.

\bibitem[{{Lique} {et~al.}(2008){Lique}, {Tobo{\l}a}, {K{\l}os}, {Feautrier},
  {Spielfiedel}, {Vincent}, {Cha{\l}asi{\'n}ski}, \& {Alexander}}]{lique08b}
{Lique}, F., {Tobo{\l}a}, R., {K{\l}os}, J., {et~al.} 2008, \aap, 478, 567.

\bibitem[{Mandelshtam \& Taylor(1995)}]{Mandelshtam-Taylor:95b}
Mandelshtam, V.~A., \& Taylor, H.~S. 1995, \jcp, 103, 2903.

\bibitem[{Mart{\'\i}nez {et~al.}(2004)Mart{\'\i}nez, Mill\'an, \&
  Gonz\'alez}]{Martinez-etal:04}
Mart{\'\i}nez, R., Mill\'an, J., \& Gonz\'alez, M. 2004, \jcp, 120,
  4705.

\bibitem[{Mart{\'\i}nez {et~al.}(2005)Mart{\'\i}nez, Sierra, \&
  Gonz\'alez}]{Martinez-etal:05}
Mart{\'\i}nez, R., Sierra, D., \& Gonz\'alez, M. 2005, \jcp, 123,
  174312.

\bibitem[{Mart{\'\i}nez {et~al.}(2006)Mart{\'\i}nez, Sierra, Gray, \&
  Gonz\'alez}]{Martinez-etal:06b}
Mart{\'\i}nez, R., Sierra, J.~D., Gray, S.~K., \& Gonz\'alez, M. 2006, \jcp, 125, 164305.

\bibitem[{{Mathis} {et~al.}(1983){Mathis}, {Mezger}, \& {Panagia}}]{Mathis1983}
{Mathis}, J.~S., {Mezger}, P.~G., \& {Panagia}, N. 1983, \aap, 128, 212.

\bibitem[{Merer {et~al.}(1975)Merer, Malm, Martin, Horani, \&
  Rostas}]{Merer-etal:75}
Merer, A.~J., Malm, D.~N., Martin, R.~W., Horani, M., \& Rostas, J. 1975, Can.
  J. Phys., 53, 251.

\bibitem[{Meuwly {et~al.}(1998)Meuwly, Meier, \& Rosmus}]{meuwly:98}
Meuwly, M., Meier, J.~P., \& Rosmus, P. 1998, \jcp, 109, 3850.

\bibitem[{Miller(1974)}]{Miller:74}
Miller, W.~H. 1974, J. Chem. Phys., 61, 1823.

\bibitem[{M{\"o}hlmann {et~al.}(1978)M{\"o}hlmann, Bhutani, de~Heer, \&
  Tsurubuchi}]{Mohlmann-etal:78}
M{\"o}hlmann, G.~R., Bhutani, K.~K., de~Heer, F.~J., \& Tsurubuchi, S. 1978,
  Chem. Phys., 31, 273.

\bibitem[{MOLPRO(2010)}]{MOLPRO}
MOLPRO. 2010, package of {\em ab initio} programs written by H.-J. Werner and
  P. J. Knowles, with contributions from R. D. Amos, A. Berning, D. L.Cooper,
  M. J. O. Deegan, A. J. Dobbyn, F. Eckert, C. Hampel, T. Leininger, R. Lindh,
  A. W. Lloyd, W. Meyer, M. E. Mura, A. Nickla\ss, P.Palmieri, K. Peterson, R.
  Pitzer, P. Pulay, G. Rauhut, M. Sch{\"u}tz, H. Stoll, A. J. Stone and T.
  Thorsteinsson.

\bibitem[{{Nagy} {et~al.}(2013){Nagy}, {Van der Tak}, {Ossenkopf}, {Gerin}, {Le
  Petit}, {Le Bourlot}, {Black}, {Goicoechea}, {Joblin}, {R{\"o}llig}, \&
  {Bergin}}]{Nagy2013}
{Nagy}, Z., {Van der Tak}, F.~F.~S., {Ossenkopf}, V., {et~al.} 2013, \aap, 550,
  A96.

\bibitem[{{Naylor} {et~al.}(2010){Naylor}, {Dartois}, {Habart}, {Abergel},
  {Baluteau}, {Jones}, {Polehampton}, {Ade}, {Anderson}, {Andr{\'e}}, {Arab},
  {Bernard}, {Blagrave}, {Bontemps}, {Boulanger}, {Cohen}, {Compi{\`e}gne},
  {Cox}, {Davis}, {Emery}, {Fulton}, {Gry}, {Huang}, {Joblin}, {Kirk},
  {Lagache}, {Lim}, {Madden}, {Makiwa}, {Martin}, {Miville-Desch{\^e}nes},
  {Molinari}, {Moseley}, {Motte}, {Okumura}, {Pinheiro Gon{\c c}alves},
  {Rod{\'o}n}, {Russeil}, {Saraceno}, {Sidher}, {Spencer}, {Swinyard},
  {Ward-Thompson}, {White}, \& {Zavagno}}]{Naylor2010}
{Naylor}, D.~A., {Dartois}, E., {Habart}, E., {et~al.} 2010, \aap, 518, L117.

\bibitem[{Neufeld {et~al.}(2010)Neufeld, {Goicoechea, J. R.}, {Sonnentrucker,
  P.}, {Black, J. H.}, {Pearson, J.}, {Yu, S.}, {Phillips, T. G.}, {Lis, D.
  C.}, {De Luca, M.}, {Herbst, E.}, {Rimmer, P.}, {Gerin, M.}, {Bell, T. A.},
  {Boulanger, F.}, {Cernicharo, J.}, {Coutens, A.}, {Dartois, E.},
  {Kazmierczak, M.}, {Encrenaz, P.}, {Falgarone, E.}, {Geballe, T. R.},
  {Giesen, T.}, {Godard, B.}, {Goldsmith, P. F.}, {Gry, C.}, {Gupta, H.},
  {Hennebelle, P.}, {Hily-Blant, P.}, {Joblin, C.}, {Kołos, R.}, {Krełowski,
  J.}, {Mart{\'\i}n-Pintado, J.}, {Menten, K. M.}, {Monje, R.}, {Mookerjea,
  B.}, {Perault, M.}, {Persson, C.}, {Plume, R.}, {Salez, M.}, {Schlemmer, S.},
  {Schmidt, M.}, {Stutzki, J.}, {Teyssier, D.}, {Vastel, C.}, {Cros, A.},
  {Klein, K.}, {Lorenzani, A.}, {Philipp, S.}, {Samoska, L. A.}, {Shipman, R.},
  {Tielens, A. G. G. M.}, {Szczerba, R.}, \& {Zmuidzinas,
  J.}}]{Neufeld-etal:10}
Neufeld, D.~A., {Goicoechea, J. R.}, {Sonnentrucker, P.}, {et~al.} 2010,
  \aap, 521, L10.

\bibitem[{{Neufeld} {et~al.}(2011){Neufeld}, {Gonz{\'a}lez-Alfonso}, {Melnick},
  {Szczerba}, {Schmidt}, {Decin}, {Alcolea}, {de Koter}, {Sch{\"o}ier},
  {Bujarrabal}, {Cernicharo}, {Dominik}, {Justtanont}, {Marston}, {Menten},
  {Olofsson}, {Planesas}, {Teyssier}, \& {Waters}}]{Neufeld2011}
{Neufeld}, D.~A., {Gonz{\'a}lez-Alfonso}, E., {Melnick}, G., {et~al.} 2011,
  \apjl, 727, L29.

\bibitem[{Neuhauser(1994)}]{Neuhauser:94}
Neuhauser, D. 1994, \jcp, 100, 9272.

\bibitem[{Pilleri {et~al.}(2014)Pilleri, {Fuente, A.}, {Gerin, M.},
  {Cernicharo, J.}, {Goicoechea, J. R.}, {Ossenkopf, V.}, {Joblin, C.},
  {Gonz\'alez-Garc{\'\i}a, M.}, {Trevi\~no-Morales, S. P.}, {S\'anchez-Monge,
  Ã.}, {Pety, J.}, {Berné, O.}, \& {Kramer, C.}}]{Pilleri-etal:14}
Pilleri, P., {Fuente, A.}, {Gerin, M.}, {et~al.} 2014, \aap, 561, A69.

\bibitem[{Porras {et~al.}(2014)Porras, Federman, Welty, \&
  Ritchey}]{Porras-etal:14}
Porras, A.~J., Federman, S.~R., Welty, D.~E., \& Ritchey, A.~M. 2014, \apjl, 781, L8.

\bibitem[{{Roueff} \& {Lique}(2013)}]{Roueff:13}
{Roueff}, E., \& {Lique}, F. 2013, Chemical Reviews, 113, 8906.

\bibitem[{{Sch{\"o}ier} {et~al.}(2005){Sch{\"o}ier}, {van der Tak}, {van
  Dishoeck}, \& {Black}}]{schoier:05}
{Sch{\"o}ier}, F.~L., {van der Tak}, F.~F.~S., {van Dishoeck}, E.~F., \&
  {Black}, J.~H. 2005, \aap, 432, 369.

\bibitem[{Skouteris {et~al.}(2000)Skouteris, Castillo, \&
  Manolopoulos}]{Skouteris-etal:00}
Skouteris, D., Castillo, J., \& Manolopoulos, D.~E. 2000, Comp. Phys. Commun.,
  133, 128, 128

\bibitem[{Smith {et~al.}(1979)Smith, Malik, \& Secrest}]{smith:79}
Smith, L.~N., Malik, D.~J., \& Secrest, D. 1979, \jcp, 71, 4502.

\bibitem[{{Spoon} {et~al.}(2013){Spoon}, {Farrah}, {Lebouteiller},
  {Gonz{\'a}lez-Alfonso}, {Bernard-Salas}, {Urrutia}, {Rigopoulou},
  {Westmoquette}, {Smith}, {Afonso}, {Pearson}, {Cormier}, {Efstathiou},
  {Borys}, {Verma}, {Etxaluze}, \& {Clements}}]{Spoon2013}
{Spoon}, H.~W.~W., {Farrah}, D., {Lebouteiller}, V., {et~al.} 2013, \apj, 775,
  127.

\bibitem[{{Tobo{\l}a} {et~al.}(2011){Tobo{\l}a}, {Dumouchel}, {K{\l}os}, \&
  {Lique}}]{Tobola:11}
{Tobo{\l}a}, R., {Dumouchel}, F., {K{\l}os}, J., \& {Lique}, F. 2011, \jcp,
  134, 024305.

\bibitem[{{Troutman} {et~al.}(2011){Troutman}, {Hinkle}, {Najita}, {Rettig}, \&
  {Brittain}}]{Troutman2011}
{Troutman}, M.~R., {Hinkle}, K.~H., {Najita}, J.~R., {Rettig}, T.~W., \&
  {Brittain}, S.~D. 2011, \apj, 738, 12.

\bibitem[{{van der Tak} {et~al.}(2007){van der Tak}, {Black}, {Sch{\"o}ier},
  {Jansen}, \& {van Dishoeck}}]{van-der-Tak2007}
{van der Tak}, F.~F.~S., {Black}, J.~H., {Sch{\"o}ier}, F.~L., {Jansen}, D.~J.,
  \& {van Dishoeck}, E.~F. 2007, \aap, 468, 627.

\bibitem[{van~der Tak {et~al.}(2013)van~der Tak, Nagy, Ossenkopf, Makai, Black,
  Faure, Gerin, \& Bergin}]{vanderTak-etal:13}
van~der Tak, F. F.~S., Nagy, Z., Ossenkopf, V., {et~al.} 2013, \aap, 560, A95.

\bibitem[{{van der Werf} {et~al.}(2010){van der Werf}, {Isaak}, {Meijerink},
  {Spaans}, {Rykala}, {Fulton}, {Loenen}, {Walter}, {Wei{\ss}}, {Armus},
  {Fischer}, {Israel}, {Harris}, {Veilleux}, {Henkel}, {Savini}, {Lord},
  {Smith}, {Gonz{\'a}lez-Alfonso}, {Naylor}, {Aalto}, {Charmandaris}, {Dasyra},
  {Evans}, {Gao}, {Greve}, {G{\"u}sten}, {Kramer}, {Mart{\'{\i}}n-Pintado},
  {Mazzarella}, {Papadopoulos}, {Sanders}, {Spinoglio}, {Stacey}, {Vlahakis},
  {Wiedner}, \& {Xilouris}}]{van-der-Werf2010}
{van der Werf}, P.~P., {Isaak}, K.~G., {Meijerink}, R., {et~al.} 2010, \aap,
  518, L42.

\bibitem[{{van Dishoeck} {et~al.}(2011){van Dishoeck}, {Kristensen}, {Benz},
  {Bergin}, {Caselli}, {Cernicharo}, {Herpin}, {Hogerheijde}, {Johnstone},
  {Liseau}, {Nisini}, {Shipman}, {Tafalla}, {van der Tak}, {Wyrowski},
  {Aikawa}, {Bachiller}, {Baudry}, {Benedettini}, {Bjerkeli}, {Blake},
  {Bontemps}, {Braine}, {Brinch}, {Bruderer}, {Chavarr{\'{\i}}a}, {Codella},
  {Daniel}, {de Graauw}, {Deul}, {di Giorgio}, {Dominik}, {Doty}, {Dubernet},
  {Encrenaz}, {Feuchtgruber}, {Fich}, {Frieswijk}, {Fuente}, {Giannini},
  {Goicoechea}, {Helmich}, {Herczeg}, {Jacq}, {J{\o}rgensen}, {Karska},
  {Kaufman}, {Keto}, {Larsson}, {Lefloch}, {Lis}, {Marseille}, {McCoey},
  {Melnick}, {Neufeld}, {Olberg}, {Pagani}, {Pani{\'c}}, {Parise}, {Pearson},
  {Plume}, {Risacher}, {Salter}, {Santiago-Garc{\'{\i}}a}, {Saraceno},
  {St{\"a}uber}, {van Kempen}, {Visser}, {Viti}, {Walmsley}, {Wampfler}, \&
  {Y{\i}ld{\i}z}}]{van-Dishoeck2011}
{van Dishoeck}, E.~F., {Kristensen}, L.~E., {Benz}, A.~O., {et~al.} 2011,
  \pasp, 123, 138.

\bibitem[{Whiting \& Nicholls(1974)}]{Whiting-Nicholls:74}
Whiting, E.~E., \& Nicholls, R.~W. 1974, Astrophys. J. Supp. Series, 27, 1, 1

\bibitem[{Whiting {et~al.}(1980)Whiting, Schadee, Taum, Hougen, \&
  Nicholls}]{Whiting-etal:80}
Whiting, E.~E., Schadee, A., Taum, J.~B., Hougen, J.~T., \& Nicholls, R.~W.
  1980, J. Mol. Spect., 80, 249.

\bibitem[{{Wyrowski} {et~al.}(2010){Wyrowski}, {van der Tak}, {Herpin},
  {Baudry}, {Bontemps}, {Chavarria}, {Frieswijk}, {Jacq}, {Marseille},
  {Shipman}, {van Dishoeck}, {Benz}, {Caselli}, {Hogerheijde}, {Johnstone},
  {Liseau}, {Bachiller}, {Benedettini}, {Bergin}, {Bjerkeli}, {Blake},
  {Braine}, {Bruderer}, {Cernicharo}, {Codella}, {Daniel}, {di Giorgio},
  {Dominik}, {Doty}, {Encrenaz}, {Fich}, {Fuente}, {Giannini}, {Goicoechea},
  {de Graauw}, {Helmich}, {Herczeg}, {J{\o}rgensen}, {Kristensen}, {Larsson},
  {Lis}, {McCoey}, {Melnick}, {Nisini}, {Olberg}, {Parise}, {Pearson}, {Plume},
  {Risacher}, {Santiago}, {Saraceno}, {Tafalla}, {van Kempen}, {Visser},
  {Wampfler}, {Y{\i}ld{\i}z}, {Black}, {Falgarone}, {Gerin}, {Roelfsema},
  {Dieleman}, {Beintema}, {de Jonge}, {Whyborn}, {Stutzki}, \&
  {Ossenkopf}}]{Wyrowski2010}
{Wyrowski}, F., {van der Tak}, F., {Herpin}, F., {et~al.} 2010, \aap, 521,
  L34+.

\bibitem[{Xu {et~al.}(2012)Xu, Li, Lv, Zhai, Duan, \& Zhang}]{Xu-etal:12}
Xu, W., Li, W., Lv, S., {et~al.} 2012, J. Phys. Chem. A, 116, 10882.

\bibitem[{Zanchet {et~al.}(2013{\natexlab{a}})Zanchet, Ag\'undez, Herrero,
  Aguado, \& Roncero}]{Zanchet-etal:13b}
Zanchet, A., Ag\'undez, M., Herrero, V.~J., Aguado, A., \& Roncero, O.
  2013{\natexlab{a}}, AJ, 146, 125.

\bibitem[{Zanchet {et~al.}(2013{\natexlab{b}})Zanchet, Godard, Bulut, Roncero,
  Halvick, \& Cernicharo}]{Zanchet-etal:13}
Zanchet, A., Godard, B., Bulut, N., {et~al.} 2013{\natexlab{b}}, ApJ, 766, 80.

\bibitem[{Zanchet {et~al.}(2009)Zanchet, Roncero, Gonz{\'a}lez-Lezana,
  Rodr{\'\i}guez-L{\'o}pez, Aguado, Sanz-Sanz, \&
  G{\'o}mez-Carrasco}]{Zanchet-etal:09b}
Zanchet, A., Roncero, O., Gonz{\'a}lez-Lezana, T., {et~al.} 2009, J. Phys.
  Chem. A, 113, 14488.

\bibitem[{Zare(1988)}]{Zare-book}
Zare, R. 1988, Angular Momentum,  John Wiley and Sons, Inc.

\bibitem[{Zhang \& Zhang(1994)}]{Zhang-Zhang:94a}
Zhang, D.~H., \& Zhang, J. Z.~H. 1994, \jcp, 101, 3672.

\end{thebibliography}

\end{document}